\documentclass[a4paper,10pt]{article}
\pdfoutput=1 

\usepackage{jheppub} 

\usepackage[T1]{fontenc} 
\usepackage{lmodern} 

\usepackage{xcolor}
\usepackage{mathrsfs,wasysym}
\usepackage{booktabs}
\usepackage{slashed}
\usepackage{cancel}

\newcommand{\kt}{k_{\rm t}}

\newcommand{\MSbar}{\overline{\text{MS}}}

\newcommand{\as}{\alpha_s}

\newcommand{\asb}{\bar\alpha_s}

\newcommand{\Ord}{\mathcal{O}}

\newcommand{\Mell}{\mathcal{M}}
\newcommand{\gammaa}{\hat\gamma}
\newcommand{\gammat}{\tilde\gamma}

\let\originalleft\left
\let\originalright\right
\renewcommand{\left}{\mathopen{}\mathclose\bgroup\originalleft}
\renewcommand{\right}{\aftergroup\egroup\originalright}

\newcommand{\sqmatr}[4]{\left(
    \begin{array}[c]{cc}
      #1 \;&\; #2 \\ #3 \;&\; #4
    \end{array}
  \right)}

\newcommand{\bc}{\chi^\sigma}

\def\beq{\begin{equation}}  
\def\eeq{\end{equation}}
\def\({\left(}
\def\){\right)}
\def\[{\left[}
\def\]{\right]}

\let\oldsubsection\subsection
\renewcommand\subsection[2][\subsectiontoc]{%
  \def\subsectiontoc{#2}%
  \oldsubsection[#1]{\boldmath #2}%
}

\let\oldsubsubsection\subsubsection
\renewcommand\subsubsection[2][\subsubsectiontoc]{%
  \def\subsubsectiontoc{#2}%
  \oldsubsubsection[#1]{\boldmath #2}%
}

\allowdisplaybreaks

\title{\boldmath Four-loop splitting functions at small $x$}

\author[a]{Marco Bonvini}
\affiliation[a]{INFN, Sezione di Roma 1,\\ Piazzale Aldo Moro~5, 00185 Roma, Italy}
\author[b]{and Simone Marzani}
\affiliation[b]{Dipartimento di Fisica, Universit\`a di Genova and INFN, Sezione di Genova,\\ Via Dodecaneso 33, 16146, Italy}
\preprint{}

\emailAdd{marco.bonvini@roma1.infn.it}
\emailAdd{simone.marzani@ge.infn.it}

\abstract{%
We consider the expansion of small-$x$ resummed DGLAP splitting functions
at next-to-leading logarithmic (NLL) accuracy to four-loop order,
namely next-to-next-to-next-to-leading order (N$^3$LO).
From this, we extract the exact LL and NLL small-$x$ contributions to the yet unknown N$^3$LO splitting functions,
both in the standard $\MSbar$ scheme and in the $Q_0\MSbar$ scheme usually considered in small-$x$ literature.
We show that the impact of unknown subleading logarithmic contributions (NNLL and beyond) at N$^3$LO is significant,
thus motivating future work towards their computation.
Our results will be also needed in future to match NLL resummation to N$^3$LO evolution.
In turn, we propose an improved implementation of the small-$x$ resummation and therefore release
a new version of the resummation code (\texttt{HELL~3.0}) which contains these changes.
}

\begin{document}

\maketitle


\section{Introduction}

The data thus far collected by the experiments at the CERN Large
Hadron Collider (LHC) have reaffirmed the Standard Model as a remarkably
successful theory of fundamental particles and their interactions.
Thus, in absence of striking signature of new physics phenomena, the
theoretical community is compelled to perform calculations with ever
smaller uncertainties so that predictions with ever increasing
accuracy and precision can be compared to data of outstanding quality,
thereby exposing subtle differences and discrepancies that may reveal
the presence of physics beyond the Standard Model.

In the context of strong interactions, accuracy is usually achieved by
computing predictions that include an increasing number of terms of
the perturbative expansion in the strong coupling $\as$ (henceforth,
the fixed-order expansion).
Leading-order (LO) cross-sections in QCD can be computed for an
essentially arbitrary number of external particles. Automation has
been achieved in recent years also for NLO calculations and an
increasing number of NNLO calculations is now available in computer
programs.
Moreover, for hadron-collider processes with simple topologies, recent
milestone calculations have achieved N$^3$LO
accuracy~\cite{Anastasiou:2015ema,Dreyer:2016oyx}.
This is particularly important because
the main production channel of the Higgs boson, i.e.\ gluon-gluon
fusion~\cite{Anzai:2015wma,Anastasiou:2016cez,Mistlberger:2018etf}, falls under this category.
Furthermore, precise theoretical predictions for LHC processes also require
precise and reliable parton distribution functions (PDFs). 
In particular, the lack of a N$^3$LO determination of PDFs is an important source
of uncertainty on the Higgs cross-section~\cite{deFlorian:2016spz}.
Although a global determination of such PDF set cannot be foreseen in
the near future, several ingredients are either already available, or
focus of current research. For instance, deep-inelastic scattering
(DIS) coefficient functions with massless quarks have been known at
three loops for a long time~\cite{Vermaseren:2005qc}, and a lot of
progress has been done in the context of heavy
quarks, e.g.\ \cite{Ablinger:2014nga,Behring:2014eya,Ablinger:2017hst,Ablinger:2017xml,Ablinger:2017err}.
Another important ingredient of this rather ambitious task is the determination of the DGLAP kernels, which
control the scale dependence of the PDFs, at N$^3$LO. Recent progress with four-loop splitting
functions~\cite{Davies:2016jie,Moch:2017uml} suggests that this calculation could be completed rather soon.

A complementary approach to the fixed-order expansion consists of exploiting all-order resummation.
In the context of PDF determination, small-$x$ (or high-energy) resummation is of particular relevance. 
Small-$x$ resummation of DGLAP evolution is known to next-to-leading logarithmic accuracy (NLL)
and it is based on the BFKL equation~\cite{Lipatov:1976zz,Fadin:1975cb,Kuraev:1976ge,Kuraev:1977fs,Balitsky:1978ic,Fadin:1998py}.
However, the proper inclusion of LL and NLL corrections is far from trivial,
due to the perturbative instability of the BFKL evolution kernel.
This problem has been tackled by more than one group in the 1990s,
see for instance Refs.~\cite{Salam:1998tj,Ciafaloni:1999yw,Ciafaloni:2003rd,Ciafaloni:2007gf},
Refs.~\cite{Ball:1995vc,Ball:1997vf,Altarelli:2001ji,Altarelli:2003hk,Altarelli:2005ni,Altarelli:2008aj}
and Refs.~\cite{Thorne:1999sg,Thorne:1999rb,Thorne:2001nr,White:2006yh},
and resulted in resummed anomalous dimensions for PDF evolution.
These techniques have recently been applied in the context of PDF determination
in Refs.~\cite{Ball:2017otu,Abdolmaleki:2018jln}, where small-$x$ resummation at NLL accuracy
has been included in the PDF evolution and in the computation of DIS structure functions
through the public code \texttt{HELL}~\cite{Bonvini:2016wki,Bonvini:2017ogt}.

It has been found that small-$x$ resummation stabilises the perturbative behaviour of both evolution kernels and partonic coefficient functions, thereby improving the description of structure-function data at low $x$.
In particular, it is well-known that potentially large logarithms at small-$x$ are absent
in NLO splitting functions due to accidental cancellations, while they start to contribute at NNLO.
As a consequence, PDFs determined with NNLO theory improved by NLL small-$x$ resummation differ rather significantly from the ones determined with NNLO alone.
Furthermore, while at NNLO the most singular term in the gluon splitting 
function is of order $\frac{\as^3}{x}\log\frac{1}{x}$ (the term with 
two logarithms being again accidentally zero), at 
N$^3$LO the most singular term is of order 
$\frac{\as^4}{x}\log^3\frac{1}{x}$. 
Hence, the aforementioned instability at low $x$ is very likely to become 
rather worse at N$^3$LO. Small-$x$ resummation would then be 
mandatory for improved precision.
We note that in order to resum all small-$x$ logarithms that appear at N$^3$LO,
one would have to consider NNLL resummation, which would be based on the three-loop BFKL kernel,
which despite a lot of recent progress~\cite
{Marzani:2007gk,DelDuca:2008jg,Bret:2011xm,DelDuca:2011ae,DelDuca:2014cya,Caron-Huot:2016tzz,Caron-Huot:2017fxr}
is not yet fully known.

In this work we examine in some detail the fixed-order expansion of the NLL
resummed splitting functions up to four loops.
This exercise is interesting for several reasons.
First, it enables us to predict the coefficients of the leading and next-to-leading
small-$x$ contributions to the yet-unknown N$^3$LO splitting functions,
thus offering either a strong check or a way of complementing the fixed-order result
at small $x$.
Second, because the resummation also includes subleading effects,
mostly related to the running of the strong coupling,
we are able to assess the impact of unknown NNLL (or higher) contributions on the four-loop result.  
Third, although we predict that N$^3$LO splitting functions will be unstable at small-$x$
(much more than the NNLO ones),
their inclusion will be most likely beneficial at moderate and large $x$,
and therefore we conclude that the most reliable result in future will be obtained
by using N$^3$LO evolution provided it is supplemented by small-$x$ resummation.
The expansion of the resummation to $\Ord(\as^4)$ presented here is also a crucial ingredient
for the N$^3$LO+NLL matching procedure.
Finally, by explicitly studying the behaviour of subleading contributions up to forth order in perturbation theory, we are able to identify a potential source of instability in our previous implementation of the resummation. We propose here an improved way of dealing with this class of subleading contributions and consequently we release a new version of the resummation code \texttt{HELL~3.0}, where these changes are implemented.

\section{\boldmath DGLAP evolution at small $x$}
\label{sec:DGLAP}

In this section we summarise small-$x$ resummation of the DGLAP splitting functions. 
Small-$x$ logarithms appear in the singlet sector and we have therefore to consider a $2\times2$
evolution matrix that couples together the quark singlet and the gluon. 
Currently, small-$x$ resummation is known to NLL and we find convenient to express resummed and matched results as
\beq \label{eq:match}
P_{ij}^\text{N$^k$LO+NLL}(x,\as) = P_{ij}^\text{N$^k$LO}(x,\as) + \Delta_{k+1}P_{ij}^{\rm NLL}(x,\as),
\qquad i,j=g,q,
\eeq
where $P_{ij}^\text{N$^k$LO}$ are the $(k+1)$-loop splitting functions and $\Delta_{k+1}P_{ij}^{\rm NLL}$
represent the resummed predictions $P_{ij}^{\rm NLL}$ minus their expansion up to order $\as^{k+1}$, namely
\beq\label{eq:DeltaPdef}
\Delta_{k+1}P_{ij}^{\rm NLL}(x,\as) = P_{ij}^{\rm NLL}(x,\as) - \sum_{j=0}^k \as^{j+1} P_{ij}^{{\rm NLL}(j)}(x),
\eeq
where $P_{ij}^{{\rm NLL}(j)}(x)$ is the $\Ord(\as^{j+1})$ contribution to $P_{ij}^{\rm NLL}(x,\as)$.
Eq.~\eqref{eq:match} is valid, in principle, for any value of $k$.
Matching of the resummation to NNLO ($k=2$ in the above notation) was achieved in Ref.~\cite{Bonvini:2017ogt} and later applied in Refs.~\cite{Ball:2017otu, Abdolmaleki:2018jln} for PDF determination.
In this work we instead focus on the matching to the next perturbative order, namely N$^3$LO ($k=3$).
We note, however, that in order to really improve the quality of the result, one should also increase the logarithmic accuracy of the resummation contribution so that no potentially large logarithm is left unresummed. Therefore, one would like to reach at least N$^3$LO+NNLL: we will leave this rather ambitious goal to future work.

Small-$x$ resummation of DGLAP evolution is usually performed in a conjugate (Mellin) space. Therefore, we define the entries of the anomalous dimension matrix in the singlet sectors as
\beq \label{eq:gamma_def}
\gamma_{ij}(N,\as) = \int_0^1 dx\, x^N P_{ij}(x,\as).
\eeq
In this non-standard notation, usually adopted in the small-$x$ resummation literature,
the leading small-$x$ logarithms of the form $\frac1x\log^k\frac1x$ are mapped into poles in $N=0$.

\subsection{Brief recap of small-$x$ resummation of DGLAP evolution}

We now recall how the resummation of DGLAP splitting functions is constructed, mainly
following Ref.~\cite{Bonvini:2017ogt}.
First, one considers the plus eigenvalue $\gamma_+(N,\as)$ of the singlet anomalous dimension matrix Eq.~\eqref{eq:gamma_def}.
This is resummed by first exploiting the duality between DGLAP evolution and BFKL evolution,
and then supplementing the result by the resummation of a class of subleading contributions
originating from the running of the strong coupling.
Additionally, requiring the symmetry of the resummed BFKL kernel and imposing momentum conservation
leads to perturbatively stable results.
Since the knowledge of the BFKL kernel at N$^k$LO allows the resummation of $\gamma_+$ at N$^k$LL,
at the moment we can only reach NLL accuracy, $\gamma_+^{\rm NLL}$.

Once the eigenvalue $\gamma_+$ is resummed, one proceeds with the resummation of $\gamma_{qg}$.
Its all-order behaviour at NLL is described by the equation~\cite{Catani:1993rn,Catani:1994sq}
\beq\label{eq:gammaqgdef}
\gamma_{qg}^{\rm NLL}(N,\as) = \as\sum_{k\geq0} h_k \[\gamma_+^k(N,\as)\],
\eeq
where the square-bracket notation is defined by the recursion~\cite{Ball:2007ra,Altarelli:2008aj}
\beq\label{eq:[]def}
\[\gamma_+^{k+1}(N,\as)\] = \Big(\gamma_+(N,\as) - k\, r(N,\as)\Big) \[\gamma_+^k(N,\as)\],
\eeq
with
\beq\label{eq:rdef}
r(N,\as) = \as^2\beta_0\frac{d}{d\as}\log\big(\gamma_+(N,\as)\big).
\eeq
Note that in Refs.~\cite{Bonvini:2016wki,Bonvini:2017ogt} a variant of the resummation where
$r(N,\as)\to\as\beta_0$, which corresponds to a limit in which $\gamma_+$ is assumed to be proportional to $\as$,
was used to infer an uncertainty on $\gamma_{qg}$.
Because only a finite number of coefficients $h_k$ are known, the implementation of the resummation,
described in Ref.~\cite{Bonvini:2016wki}, is only approximate.
However, for not-too-large values of $\as$, the implementation is numerically stable and reliable.

Simple power-counting at small-$N$ shows that the quark anomalous dimension $\gamma_{qg}$ starts at NLL.
Therefore, at this accuracy we have some freedom in how we choose the logarithmic accuracy
of $\gamma_+$ appearing in Eq.~\eqref{eq:gammaqgdef}.
In Ref.~\cite{Bonvini:2016wki} a dedicated anomalous dimension, denoted LL$^\prime$,
was constructed specifically for this purpose.
This LL$^\prime$ anomalous dimension is essentially a LL anomalous dimension,
but its singular structure, which at resummed level is encoded in the position of the rightmost pole in $N$ space,
is taken from the NLL result.
The reason for using this hybrid object can be summarised as follows:
\begin{itemize}
\item on the one hand, it is preferable to use the $\gamma_+^\text{NLL}$ anomalous dimension in order to avoid singularity mismatches between different entries of the anomalous dimension matrix;
\item on the other hand, since using $\gamma_+$ at LL in Eq.~\eqref{eq:gammaqgdef} is formally sufficient to achieve NLL accuracy in $\gamma_{qg}$, it was convenient from a numerical point of view to use as much of the LL result as possible,
  because of its better numerical stability.
\end{itemize}
However, recently in Ref.~\cite{Bonvini:2017ogt} various improvements in the construction of the resummed anomalous dimensions
have been proposed and implemented in the numerical code \texttt{HELL}.
With these developments, the computation of the full NLL anomalous dimension is faster and much more stable and reliable,
and it is therefore now possible to either use the hybrid LL$^\prime$ result or the full NLL one in Eq.~\eqref{eq:gammaqgdef}.
In particular, the latter choice corresponds to the original approach of Ref.~\cite{Altarelli:2008aj}.
We will explore the effects of both options in the following.

All the other entries of the singlet anomalous dimension matrix can be derived from the plus eigenvalue and $\gamma_{qg}$.
In particular, the results can be written in a rather simple form if we consider
the resummed contributions $\Delta_k \gamma_{ij}$, defined according to Eq.~\eqref{eq:DeltaPdef}:
\begin{subequations}\label{eq:Delta4gamma}
\begin{align}
\Delta_k\gamma_{gg}^{\rm NLL}(N,\as) &= \Delta_k\gamma_+^{\rm NLL}(N,\as) -\frac{C_F}{C_A}\Delta_k\gamma_{qg}^{\rm NLL}(N,\as), \\
\Delta_k\gamma_{gq}^{\rm NLL}(N,\as) &= \frac{C_F}{C_A}\Delta_k\gamma_{gg}^{\rm NLL}(N,\as), \\
\Delta_k\gamma_{qq}^{\rm NLL}(N,\as) &= \frac{C_F}{C_A}\Delta_k\gamma_{qg}^{\rm NLL}(N,\as).
\end{align}
\end{subequations}
We recall that these 
relations are able to predict only the LL part of $\gamma_{gq}$,
while we do not have enough knowledge to predict its NLL part, which is then only approximate in the equation above.
Having all the anomalous dimensions in the singlet sector,
we can construct the splitting functions by Mellin inversion.\footnote
{In phenomenological applications, a damping at large $x$ is added to make the transition
  from the small-$x$ region (where the resummation is relevant)
  to the large-$x$ region (where the fixed-order description is appropriate) as smooth as possible,
  and finally momentum conservation is reimposed.
  These details have been described in Ref.~\cite{Bonvini:2017ogt} and are not repeated here.}

\subsection{Perturbative expansion of the resummation}

We now consider the perturbative expansion of the resummed result presented in the previous section.
The goal is twofold. On the one hand, the expansion of the resummation is needed to construct
the resummed contributions $\Delta_k\gamma_{ij}^{\rm NLL}$, $\Delta_kP_{ij}^{\rm NLL}$, Eq.~\eqref{eq:DeltaPdef},
namely for the matching of the resummed result to fixed order.
On the other hand, the $\as$ expansion of the resummed results provides a prediction
for the small-$x$ behaviour of the fixed-order splitting functions.

In Ref.~\cite{Bonvini:2017ogt} we have already determined the expansion of the NLL resummed splitting functions
to $\Ord(\as^3)$, which was needed to match resummation to NNLO.
In that case, there was no point in using the result of the expansion to predict the NNLO behaviour at small-$x$,
as the three-loop splitting functions are known~\cite{Vogt:2004mw,Moch:2004pa}.
In this work we push the expansion to one extra order, $\Ord(\as^4)$.
These results would be needed to match resummation to N$^3$LO, and specifically to construct N$^3$LO+NLL resummed results.
The four-loop splitting functions, however, are not yet fully known~\cite{Davies:2016jie,Moch:2017uml}.
Therefore, at the moment our results can be used to construct approximations, valid at small $x$,
of the unknown N$^3$LO splitting functions, or simply to supplement the ongoing computation
with the knowledge of the exact small-$x$ behaviour.
In future, when the four-loop splitting functions will be computed, it will also serve as a cross check.

In order to obtain the expansion of the resummed entries of the anomalous dimension matrix, 
we have to expand both $\gamma_+^{\rm NLL}$ and $\gamma_{qg}^\text{NLL}$ to the desired accuracy;
the other anomalous dimensions are recovered using Eqs.~\eqref{eq:Delta4gamma}.
Let us first introduce a generic notation for the expansion of the plus anomalous dimension,
\beq\label{eq:gamma+exp}
\gamma_+(N,\as) = \as \gamma_0 + \as^2 \gamma_1 + \as^3 \gamma_2 + \as^4 \gamma_3 + \Ord(\as^5),
\eeq
which is valid both for the NLL anomalous dimension and for the auxiliary LL$^\prime$ one.
The expansion of the $qg$ anomalous dimension, according to Eqs.~\eqref{eq:gammaqgdef}, \eqref{eq:[]def} and \eqref{eq:rdef},
is given by
\begin{align}\label{eq:gammaqg}
\gamma_{qg}^{\rm NLL}(N,\as)
&= \as h_0 + \as^2 h_1 \gamma_0
+ \as^3 \[h_2 \gamma_0\(\gamma_0 -\beta_0\) + h_1 \gamma_1\] \nonumber\\
&+ \as^4 \[h_3 \gamma_0 \(\gamma_0 -\beta_0\) \(\gamma_0 -2\beta_0\)+ h_2 \gamma_1(N) \(2\gamma_0 -T\beta_0\) + h_1 \gamma_2\] \nonumber\\
&+ \Ord(\as^5),
\end{align}
where the anomalous dimensions $\gamma_{0,1,2}$ are the coefficients Eq.~\eqref{eq:gamma+exp}
of the expansion of either $\gamma_+^\text{NLL}$ or $\gamma_+^{\text{LL}'}$, 
depending on the choice adopted in Eq.~\eqref{eq:gammaqgdef}.
Note that the expansion Eq.~\eqref{eq:gammaqg} depends on a parameter $T$. This parameter
has been introduced to account for the two variants of the resummation of running coupling
contributions described above. In particular, when $r(N,\as)$ as given by Eq.~\eqref{eq:rdef} is used then $T=2$,
while for the variation $r(N,\as)\to\as\beta_0$ then $T=1$.
The first $h_k$ coefficients appearing in Eq.~\eqref{eq:gammaqg} are~\cite{Catani:1994sq}
\beq
h_0=\frac{n_f}{3\pi},\qquad
h_1=\frac{n_f}{3\pi}\,\frac53,\qquad
h_2=\frac{n_f}{3\pi}\,\frac{14}9,\qquad
h_3=\frac{n_f}{3\pi}\(\frac{82}{81} + 2\zeta_3\).
\eeq
Note that the knowledge of $\gamma_{qg}^{\rm NLL}$ to $\Ord(\as^4)$ requires the expansion of the plus
anomalous dimension up to $\Ord(\as^3)$, as it does not depend on $\gamma_3$.
This is due to the fact that $\gamma_{qg}$ is a pure NLL quantity,
as it is clear from the factor of $\as$ in front of Eq.~\eqref{eq:gammaqgdef}.

We thus need to compute the first four orders of $\gamma_+^{\rm NLL}$,
while we just need the first three orders of $\gamma_+^{\rm LL'}$
as it only possibly enters in the expansion of $\gamma_{qg}$.
The precise construction of these resummed anomalous dimensions was presented in detail in Ref.~\cite{Bonvini:2017ogt},
and we do not repeat it here (some details are given in App.~\ref{sec:expansion}).
We just recall that due to the actual construction of the plus eigenvalue,
which is based on the duality between DGLAP and BFKL evolutions,
the LL and LL$^\prime$ resummed anomalous dimensions automatically contain the fixed LO
anomalous dimension, while the NLL anomalous dimension contains the NLO one.\footnote
{These fixed-order anomalous dimensions used in the construction of resummation are actually approximated,
  as explained in Ref.~\cite{Bonvini:2017ogt}. This fact is however immaterial for the present discussion.}
However, the $qg$ anomalous dimension, Eq.~\eqref{eq:gammaqgdef}, requires a purely resummed anomalous dimension,
which goes to zero at large $N$, in order to avoid producing spurious large-$N$ terms
(see discussion in App.~B.2 of Ref.~\cite{Bonvini:2016wki}).
Therefore, we first define the resummed contributions
$\Delta_1\gamma_+^{\rm LL'}$ and $\Delta_2\gamma_+^{\rm NLL}$
to be the resummed results at LL$^\prime$ and NLL minus the LO and NLO anomalous dimensions, respectively
(the notation is the same as in Ref.~\cite{Bonvini:2017ogt}).
Then, we construct the purely resummed LL$^\prime$ and NLL anomalous dimensions as
\begin{subequations}
\begin{align}
\gamma_+^{\rm LL'}(N,\as) &= \as\gamma_0^{\rm LL'}(N) + \Delta_1\gamma_+^{\rm LL'}(N,\as) ,\\
\gamma_+^{\rm NLL}(N,\as) &= \as\gamma_0^{\rm NLL}(N) + \as^2\gamma_1^{\rm NLL}(N) + \Delta_2\gamma_+^{\rm NLL}(N,\as) .
\end{align}
\end{subequations}
The functions $\gamma_0^{\rm LL'}$, $\gamma_0^{\rm NLL}$ and $\gamma_1^{\rm NLL}$ are not fixed by the resummation,
and we thus have a degree of arbitrariness in how to define them.
Instead, the expansions of the resummed contributions
\begin{subequations}
\begin{align}
\Delta_1\gamma_+^{\rm LL'}(N,\as) &= \as^2\gamma_1^{\rm LL'}(N) + \as^3\gamma_2^{\rm LL'}(N) + \Ord(\as^4) ,\\
\Delta_2\gamma_+^{\rm NLL}(N,\as) &= \as^3\gamma_2^{\rm NLL}(N) + \as^4\gamma_3^{\rm NLL}(N) + \Ord(\as^5)
\end{align}
\end{subequations}
can be derived from the resummed results of Ref.~\cite{Bonvini:2017ogt}.
Their computation is presented in App.~\ref{sec:expansion}.
Part of them, specifically $\gamma_1^{\rm LL'}$ and $\gamma_2^{\rm NLL}$,
have been already computed and presented in Ref.~\cite{Bonvini:2017ogt}.
The next terms, $\gamma_2^{\rm LL'}$ and $\gamma_3^{\rm NLL}$, are reported here for the first time.

Starting from LL$^\prime$ resummation, the first order of the anomalous dimension was chosen in Ref.~\cite{Bonvini:2016wki}
to include the LL and NLL contributions of the LO anomalous dimension. The NLL term, being it a constant at this order,
is further multiplied by a function $1/(N+1)$ to make it vanish at large $N$. The expression, which we adopt also here, is
\beq
\gamma_0^{\rm LL'} = \frac{a_{11}}{N} + \frac{a_{10}}{N+1},
\eeq
where $a_{ij}$ are defined in Eq.~\eqref{eq:aij}.
The next orders, as predicted by the resummation, are given by (see App.~\ref{sec:expansion})
\begin{align}
\gamma_1^{\rm LL'} &= \beta_0\(\frac{3}{32}\kappa_0-c_0\)\(\frac1N-\frac{4N}{(N+1)^2}\) \label{eq:gamma1LLp}\\
\gamma_2^{\rm LL'} &= \frac{\lambda_2}{N^2}+\frac{\lambda_1}{N} - \(\lambda_2+\lambda_1\) \frac{4N}{(N+1)^2} \nonumber\\
                   &\quad+\(\frac{a_{11}}{N^2}+\frac{2(a_{11}+a_{10})}{(N+1)^2}\)
                     \bigg[ \frac{a_{11}a_{10}}{(N+1)^2}-\frac{a_{11}a_{10}}{4} \frac{4N}{(N+1)^2} \nonumber\\
&\qquad \qquad \qquad \qquad \qquad \qquad\; +a_{11}\(\frac{a_{11}}N+a_{10}-\frac{2(a_{11}+a_{10})N}{N+1}\)\[\psi_1(N+1)-\zeta_2\]\bigg],
\label{eq:gamma2LLp}
\end{align}
where all the coefficients are defined in App.~\ref{sec:expansion}.
Eq.~\eqref{eq:gamma2LLp} is a new result.
Note that, by construction, both $\gamma_1^{\rm LL'}$ and $\gamma_2^{\rm LL'}$ vanish in $N=1$, as the resummation is built to preserve momentum conservation.

Moving to the NLL resummation, we need to choose both the LO and NLO contributions.
We decide to adopt the same strategy as in the LL$^\prime$, i.e.\ keeping only the LL and NLL contributions from the fixed orders.
In particular, for the LO term we use exactly the same approximation used for the LL$^\prime$,
and at NLO, due to the fact that the LL term is accidentally zero, we simply have a NLL term,
\begin{align}
\gamma_0^{\rm NLL} &= \frac{a_{11}}{N} + \frac{a_{10}}{N+1}, \\
\gamma_1^{\rm NLL} &= \frac{a_{21}}{N} - \frac{2a_{21}}{N+1},
\end{align}
where $a_{21}$, the NLL coefficient of the NLO, is defined in Eq.~\eqref{eq:aij}.
In $\gamma_1^{\rm NLL}$ we have also included a subtraction term of the same form as the NLL term in $\gamma_0^{\rm NLL}$,
which restores momentum conservation, i.e.\ $\gamma_1^{\rm NLL}(1)=0$.
We have decided to add this feature as the effect is formally NNLL, and it makes it more in line with the LL$^\prime$ case where
the $\Ord(\as^2)$ term vanishes in $N=1$.
We stress that we have played with variants of both $\gamma_0^{\rm NLL}$ and $\gamma_1^{\rm NLL}$, adding momentum conservation to the first,
relaxing it in the second, varying the way it is implemented, and so on:
the effect at the level of the splitting functions is moderate.
The third and fourth coefficients are instead found expanding the resummation,
and their form is (see App.~\ref{sec:expansion})
\begin{align}
\gamma_2^{\rm NLL} &= \beta_0^2 \frac{\kappa_0}{16} \(\frac1N - \frac{4N}{(N+1)^2}\) \nonumber\\
                   &\quad+\(\frac{a_{11}}{N^2}+\frac{2(a_{11}+a_{10})}{(N+1)^2}\)
                     \bigg[ \rho + \frac{a_{21}}{N+1} + \frac{a_{11}a_{10}}{(N+1)^2} -\(\rho + \frac{a_{21}}2 + \frac{a_{11}a_{10}}4 -\beta_0a_{11}\) \frac{4N}{(N+1)^2} \nonumber\\
&\qquad \qquad \qquad \qquad \qquad \qquad\; +a_{11}\(\frac{a_{11}}N+a_{10}-\frac{2(a_{11}+a_{10})N}{N+1}+\beta_0\)\[\psi_1(N+1)-\zeta_2\]\bigg], \label{eq:gamma2NLL} \\
\gamma_3^{\rm NLL} &= \frac{\rho_3}{N^3} +\frac{\rho_2}{N^2} +\frac{\rho_1}{N} - (\rho_3+\rho_2+\rho_1) \frac{4N}{(N+1)^2} \nonumber\\
&\quad+\(\frac{a_{21}}{N^2}+\frac{2(a_{21}+a_{20})}{(N+1)^2}\)
\bigg[ \rho + \frac{a_{21}}{N+1} + \frac{a_{11}a_{10}}{(N+1)^2} -\(\rho + \frac{a_{21}}2 + \frac{a_{11}a_{10}}4 -\beta_0a_{11}\) \frac{4N}{(N+1)^2} \nonumber\\
&\qquad \qquad \qquad \qquad \qquad \qquad\; +a_{11}\(\frac{a_{11}}N+a_{10}-\frac{2(a_{11}+a_{10})N}{N+1}+\beta_0\)\[\psi_1(N+1)-\zeta_2\]\bigg]\nonumber\\
&\quad+\(\frac{a_{11}^2}{N^2}+\frac{2a_{11}(a_{11}+a_{10})}{(N+1)^2}\) \Bigg\{
\(\frac{a_{10}^2}{2}-a_{10}(a_{11}-\beta_0)-a_{11}^2\)\(\frac{2}{(N+1)^3}-\frac{N}{(N+1)^2}\)
\nonumber\\ 
&\qquad\qquad\qquad\qquad\qquad\qquad\quad
+ \(\frac{a_{21}a_{10}}{a_{11}}+a_{20}\) \(\frac{1}{(N+1)^2} - \frac{N}{(N+1)^2}\)
\nonumber\\ 
&\qquad\qquad\qquad\qquad\qquad\qquad
+\(\frac{a_{21}}N+a_{20}-\frac{2(a_{21}+a_{20})N}{N+1}\)\[\psi_1(N+1)-\zeta_2\]
\nonumber\\ 
&\qquad\qquad\qquad\qquad\qquad\quad
+\(\frac{a_{11}}N+a_{10}-\frac{2(a_{11}+a_{10})N}{N+1}\)^2\(\zeta_3-\frac12\psi_2(N+1)\)
\nonumber\\ 
&\qquad\qquad\qquad\qquad\qquad
+\(\frac{a_{11}}N+a_{10}-\frac{2(a_{11}+a_{10})N}{N+1}\)
\bigg[\(3\zeta_3-\frac12\psi_2(N+1)\)\beta_0
+ \frac{\tilde\chi_1'(0,N)}{a_{11}}
\nonumber\\ 
&\qquad\qquad\qquad\qquad\qquad\qquad\qquad\qquad\qquad\qquad\qquad\qquad\qquad\qquad
+\frac{2a_{10}}{(N+1)^3}
  + \frac{a_{21}/a_{11}}{(N+1)^2}\bigg]
\Bigg\},
\label{eq:gamma3NLL}
\end{align}
where again the various coefficients are defined in App.~\ref{sec:expansion},
and we have left implicit the function $\tilde\chi_1'(0,N)$, Eq.~\eqref{eq:chi00+}.
Eq.~\eqref{eq:gamma2NLL} was already presented in Ref.~\cite{Bonvini:2017ogt},\footnote
{Note that Eq.~\eqref{eq:gamma2NLL} differs from the analogous result of Ref.~\cite{Bonvini:2017ogt}
by the subleading (NNLL) $\beta_0$ terms appearing in the second and third lines.
Their origin is discussed in App.~\ref{sec:bug}.}
while Eq.~\eqref{eq:gamma3NLL} is a new result of this study.

Before moving further, some comments are in order.
We observe that, due to accidental zeros of the LL singularity both at NLO and NNLO,
the leading singularity at these two orders is the NLL one, namely $1/N$ and $1/N^2$, respectively.
These NLL poles are predicted correctly by NLL resummation,
so in particular $\gamma_2^{\rm NLL}$ has the exact leading singularity
($\gamma_1^{\rm NLL}$ has it by construction).
In contrast, the leading singularity of both $\gamma_1^{\rm LL'}$ and $\gamma_2^{\rm LL'}$
is predicted by the resummation and thus, being the resummation just accurate at LL, is not exact.
While the full LL$^\prime$ anomalous dimension, being an all-order result, is reliable,
each term of its perturbative expansion may not be. In particular, these two terms,
$\gamma_1^{\rm LL'}$ and $\gamma_2^{\rm LL'}$, do not contain anything of the exact result,
because they are zero at LL.
Since the impact of these two orders in the expansion of resummed splitting functions
(in particular $P_{qg}$ and $P_{qq}$) may be substantial,
we may expect that using LL$^\prime$ resummation may give rise to somewhat unreliable
resummed contributions when matched to high orders, e.g.\ NNLO or N$^3$LO.
More precisely, we may expect NLL resummation, which has the exact leading contributions at these two orders,
to lead to more reliable matched results to high orders.

These considerations suggest that the use of the NLL anomalous dimension in the construction
of $\gamma_{qg}$ (as originally suggested in Ref.~\cite{Altarelli:2008aj}) is preferred.
Thus, from now on we will adopt the NLL anomalous dimension as default ingredient
for the resummation of anomalous dimension matrix,
and possibly consider LL$^\prime$ resummation as a variant
to estimate the impact of subleading logarithmic contributions.
Both options are now available in the new \texttt{3.0} version of the \texttt{HELL} code.

\subsection{The four-loop splitting functions at small-$x$ in the $Q_0\MSbar$ scheme}
\label{sec:Q0MSbar}

The results of the previous section allow to construct all $\Delta_4\gamma_{ij}^{\rm NLL}$,
and thus by Mellin inversion all $\Delta_4P_{ij}^{\rm NLL}$, needed to match NLL resummation to N$^3$LO evolution.
While these results will be in practice of no use until
the computation of the N$^3$LO splitting functions will be completed,
it is interesting to extract from them the small-$x$ terms at LL and NLL
of the yet unknown four-loop splitting functions.
These may be also useful to construct approximate predictions
of the four-loop splitting functions while waiting for their full computation.

We remind the reader that the resummation procedure previously
described lead to all-order results which are not in the traditional
$\MSbar$ scheme, but rather in a related factorization scheme called
$Q_0\MSbar$~\cite{Catani:1993ww,Catani:1994sq,Ciafaloni:2005cg,Marzani:2007gk},
which is particularly suitable for small-$x$ resummation.
Indeed, in the $\MSbar$ scheme there are some cancellations of large small-$x$ contributions taking place between
parton evolution and coefficient functions. The $Q_0$ variant of the scheme automatically
removes these large contributions from both objects, leading to resummed predictions
which are perturbatively much more stable.
For this reason, here as well as in previous studies, we always use the $Q_0\MSbar$ scheme
for results including small-$x$ resummation.
The difference between the $\MSbar$ and $Q_0\MSbar$ factorization schemes influences
the resummation of the anomalous dimensions beyond the leading logarithmic accuracy,
as well as the resummation of the coefficient functions.
We first concentrate on $Q_0\MSbar$, while we present results for $\MSbar$ in the next section.

The small-$N$ expansion of the expansion terms of $\gamma_+^{\rm NLL}$ is given in App.~\ref{sec:expansion}.
With those results, we can construct all the other entries using Eqs.~\eqref{eq:gammaqg} and \eqref{eq:Delta4gamma}.
Denoting with $\gamma_{ij}^{(k)}$ the exact $\Ord(\as^{k+1})$ anomalous dimension, we have
\begin{subequations}\label{eq:gamma3sx}
\begin{align}
\gamma_{gg}^{(3)}(N) &= \frac1{N^4} \frac{C_A^4}{\pi^4}2\zeta_3 \nonumber\\
& + \frac1{N^3}\frac{1}{\pi^4} \bigg[
  C_A^4\(-\frac{1205}{162} + \frac{67}{36}\zeta_2 + \frac14 \zeta_2^2 - \frac{77}{6} \zeta_3\)
  + n_fC_A^3\(-\frac{233}{162} + \frac{13}{36}\zeta_2 + \zeta_3\)\nonumber\\
  &\qquad\qquad\quad+ n_fC_A^2C_F\(\frac{617}{243} - \frac{13}{18}\zeta_2 + \frac23 \zeta_3\)
\bigg] \nonumber\\
& + \Ord\(\frac1{N^2}\), \\
\gamma_{qg}^{(3)}(N) &= \frac1{N^3}\frac{C_A^3n_f}{3\pi^4}\(\frac{82}{81} + 2\zeta_3\) + \Ord\(\frac1{N^2}\).
\end{align}
\end{subequations}
The $gq$ and $qq$ anomalous dimensions are obtained by simply multiplying the $gg$ and $qg$ ones by $C_F/C_A$,
even though we stress again that only the $1/N^4$ pole of the resulting $gq$ anomalous dimension is correct.
Hence we have
\begin{subequations}\label{eq:gamma3sxquark}
\begin{align}
\gamma_{gq}^{(3)}(N) &= \frac1{N^4} \frac{C_A^3C_F}{\pi^4}2\zeta_3 + \Ord\(\frac1{N^3}\), \\
\gamma_{qq}^{(3)}(N) &= \frac1{N^3}\frac{C_A^2C_Fn_f}{3\pi^4}\(\frac{82}{81} + 2\zeta_3\) + \Ord\(\frac1{N^2}\).
\end{align}
\end{subequations}
The corresponding small-$x$ logarithms in the four-loop splitting functions are easily obtained by Mellin inversion of Eqs.~\eqref{eq:gamma3sx},
according to
\beq
\Mell^{-1}\[\frac1{N^4}\] = \frac1{6}\frac{\log^3\frac1x}{x}, \qquad
\Mell^{-1}\[\frac1{N^3}\] = \frac1{2}\frac{\log^2\frac1x}{x},
\eeq
where $\Mell^{-1}$ denotes the inverse Mellin transform. Thus, four-loop splitting functions exhibit a much stronger growth at small-$x$ than those at a previous order, which behave like $\as^3 x^{-1} \log x$.
To our knowledge, the NLL contribution to $P_{gg}$ is explicitly presented here for the first time.

\subsection{The four-loop splitting functions at small-$x$ in the $\MSbar$ scheme}
\label{sec:MSbar}

The effect of scheme change between $Q_0\MSbar$ and $\MSbar$ turns out to be of relative order $\as^3$,
and thus all the fixed-order results considered in previous studies on small $x$ resummation
happened to be identical in either scheme.
However, in this work we are considering resummation matched (or expanded) to N$^3$LO,
thus becoming sensitive to the scheme choice even at fixed order.
It is thus important to recall how the conversion is performed.
The goal of this subsection is also to provide the small-$x$ contributions
of the N$^3$LO splitting functions in the $\MSbar$ scheme,
namely in the scheme in which the full four-loop computation will likely be carried out.

A factorization scheme change is a multiplicative redefinition of the PDFs $f$ and coefficient functions $C$.
Focussing on $\MSbar$ and $Q_0\MSbar$, and considering both processes with one or two hadrons in the initial state
(i.e.\ coefficient functions with one or two flavour indices), we have
\begin{align}
f_i^{\MSbar}(N,Q^2) &= \Lambda_{ij}^{-1}(N,\as) f_j(N,Q^2), \\
C_i^{\MSbar}(N,\as) &= C_j(N,\as) \Lambda_{ji}(N,\as), \\
C_{ij}^{\MSbar}(N,\as) &= C_{kl}(N,\as) \Lambda_{ik}(N,\as) \Lambda_{jl}(N,\as),
\end{align}
where $\as=\as(Q^2)$ and we denoted with a $\MSbar$ label quantities in that scheme
and without label quantities in the $Q_0\MSbar$ scheme.
Accordingly, the anomalous dimensions change as
\beq\label{eq:gammaSC}
\gamma_{ij}^{\MSbar}(N,\as) = \Lambda_{ik}^{-1}(N,\as) \gamma_{kl}(N,\as) \Lambda_{lj}(N,\as) - \Lambda_{ik}^{-1}(N,\as) Q^2\frac{d \Lambda_{kj}(N,\as) }{dQ^2}.
\eeq
The function $\Lambda_{ij}$ is a matrix in flavour space implementing the scheme change.
As far as small-$x$ scheme changes are concerned, this matrix is trivial in its non-singlet part, so we focus only on the singlet.
Up to NLL, its form is given by~\cite{Altarelli:2008aj}\footnote
{Note that there is a typo in Ref.~\cite{Altarelli:2008aj} that we correct here.}
\beq\label{eq:Usc}
\sqmatr{\Lambda_{gg}}{\Lambda_{gq}}{\Lambda_{qg}}{\Lambda_{qq}} =
\sqmatr{R}{\frac{C_F}{C_A}(R-1)}{0}{1} + \as \sqmatr{\cdot}{\cdot}{v}{\cdot} + \text{NNLL},
\eeq
where both $R$ and $v$ are LL functions, i.e.\ functions of $\as/N$ to all orders.
The form of the LL part of the matrix is such that the scheme change has no effect on the LL part of the anomalous dimension matrix.
The three empty slots in the NLL part of the matrix have an effect only on the $\gamma_{gq}$ entry at NLL,
which is however not determined by NLL resummation (as we already stressed), and are thus of no relevance.
Furthermore, they also affect the resummation of partonic coefficient functions but this effect is beyond the accuracy currently achieved in the context of small-$x$ resummation.
Thus, at the currently available logarithmic accuracy, the three empty slots can be any LL function of $\as/N$.

The scheme-change function was calculated long ago~\cite{Catani:1994sq}
\begin{align}
R(M) &= \sqrt{\frac{-1}{M}\frac{\Gamma\(1-M\) \chi_0 \(M\)}{\Gamma\(1+M\) \chi_0^\prime \(M\)} }
\exp \left \{M \psi(1)+\int_0^{M} d c \, \frac{\psi^\prime(1)-\psi^\prime \(1-c\)}{2\psi(1)-\psi(c)-\psi(1-c)} \right\},\\
v(M) &= \frac{R(M)-1}M h(M),\label{eq:vdef}
\end{align}
where $h(M)=\sum_{k\geq0}h_kM^k$ is the function used for resumming $\gamma_{qg}$, Eq.~\eqref{eq:gammaqgdef},
$\chi_0(M)$ is the BFKL kernel at LO (Eq.~\eqref{eq:chi0} in $N=0$),
$\Gamma(M)$ and $\psi(M)$ the gamma and digamma functions respectively.
The functions $R(M)$ and $v(M)$ have to be evaluated in $M=\gamma_+(N,\as)$ in Eq.~\eqref{eq:Usc}, where $\gamma_+$ is the resummed one
(in either scheme, as only its LL part needs to be correct, and at LL the scheme change is ineffective on the anomalous dimensions).
The form of $v$, Eq.~\eqref{eq:vdef}, is such that at NLL both $\gamma_{qg}$ and $\gamma_{qq}$
are the same in $\MSbar$ and $Q_0\MSbar$.
Additionally, we have already noted that the matrix structure of $\Lambda_{ij}$ at LL is such that
is has no effect on the LL anomalous dimensions. Thus, the only anomalous dimension
which is sensitive at NLL to the scheme change is $\gamma_{gg}$, and we have
\beq
\gamma_{gg}^{\MSbar}(N,\as) = \gamma_{gg}(N,\as) + \as^4\[\beta_0 8\zeta_3 \gamma_0^3(N)+\Ord\(\frac1{N^2}\)\] + \Ord(\as^5),
\eeq
having used the expansion of the function $R$ in powers of $M$,
\beq
R(M)=1+\frac83\zeta_3M^3+\Ord(M^4).
\eeq
The scheme change contribution is entirely due to the derivative term in Eq.~\eqref{eq:gammaSC}.
Of course, also the NLL part of $\gamma_{gq}$ changes by the scheme change, but as we already repeated several times
NLL resummation is not able to predict it.
To conclude, we report the actual expansion to NLL of the $gg$ anomalous dimension in the $\MSbar$ scheme:
\begin{align}\label{eq:gamma3sxMSbar}
\gamma_{gg}^{\MSbar(3)}(N) &= \frac1{N^4} \frac{C_A^4}{\pi^4}2\zeta_3 \nonumber\\
& + \frac1{N^3}\frac{1}{\pi^4} \bigg[
  C_A^4\(-\frac{1205}{162} + \frac{67}{36}\zeta_2 + \frac14 \zeta_2^2 - \frac{11}{2} \zeta_3\)
  + n_fC_A^3\(-\frac{233}{162} + \frac{13}{36}\zeta_2 - \frac13 \zeta_3\)\nonumber\\
  &\qquad\qquad\quad+ n_fC_A^2C_F\(\frac{617}{243} - \frac{13}{18}\zeta_2 + \frac23 \zeta_3\)
\bigg] \nonumber\\
& + \Ord\(\frac1{N^2}\).
\end{align}
At this accuracy, all the other entries are identical to their $Q_0\MSbar$ counterparts given in Sect.~\ref{sec:Q0MSbar}.
To our knowledge, the NLL contributions to $P_{gg}$ in the $\MSbar$ scheme are explicitly presented here for the first time.

\section{Numerical results and discussion}
\label{sec:results}

Thus far we have presented analytical results.
We now concentrate on numerics and we illustrate the difference between the two variants of the resummation discussed above.
We also present approximate results for the four-loop splitting functions which are based on the expansion of the resummation and we critically assess the trustworthiness of this construction.

\subsection{Resummed splitting functions at NNLO+NLL}
\begin{figure}[t]
  \centering
  \includegraphics[width=0.495\textwidth,page=1]{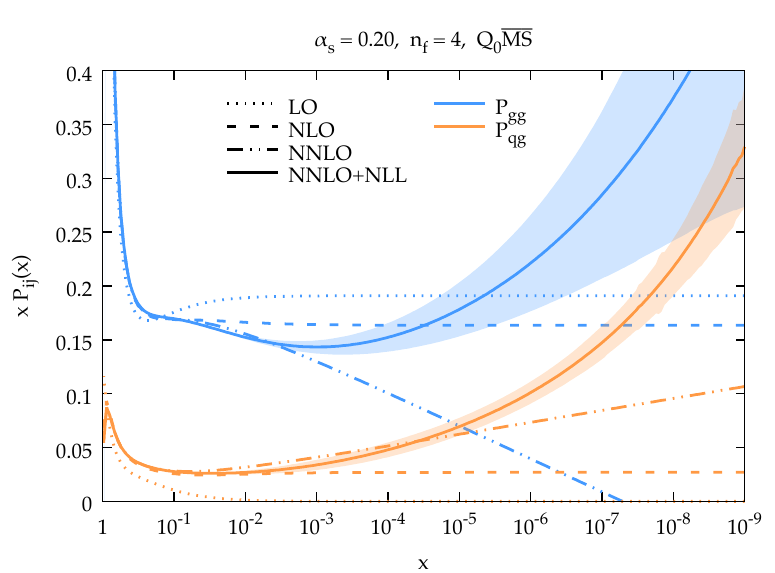}
  \includegraphics[width=0.495\textwidth,page=2]{images/plot_P_nf4_as020_mixed.pdf}
  \caption{The resummed and matched splitting functions at NNLO+NLL (solid) accuracy
    compared with the fixed-order results at LO (dotted), NLO (dashed) and NNLO (dot-dot-dashed).
    The left plot shows $P_{gg}$ (blue) and $P_{qg}$ (orange), and the right plot $P_{gq}$ (blue) and $P_{qq}$ (orange).
    The plots are for $\as=0.2$ and $n_f=4$ in the $Q_0\MSbar$ scheme.}
  \label{fig:Pres}
\end{figure}
First, we consider the four singlet splitting functions at fixed order and with resummation
using the NLL anomalous dimension, which is the new default in \texttt{HELL 3.0}.
In Fig.~\ref{fig:Pres} we show $P_{gg}$ (blue) and $P_{qg}$ (orange) in the left plot,
and $P_{gq}$ (blue) and $P_{qq}$ (orange) in the right plot,
at LO (dotted), NLO (dashed), NNLO (dot-dot-dashed) and NNLO+NLL (solid).
The resummed result is supplemented with an uncertainty band,
which aims to estimate the impact of unknown subleading logarithmic contributions.
Following Ref.~\cite{Bonvini:2017ogt}, this band is obtained by considering variations
of the way RC resummation of $\gamma_+$ is implemented and of the way the resummation of 
$\gamma_{qg}$ is performed, and summing in quadrature the two effects.\footnote
{In fact, in Ref.~\cite{Bonvini:2017ogt} we considered only the second variation for $P_{qg}$ and $P_{qq}$;
we now use a more symmetric approach and use both for all the splitting functions.}
The qualitative aspect of these results is the same of those obtained with the \texttt{HELL 2.0}
settings of Ref.~\cite{Bonvini:2017ogt}, i.e.\
using the LL$^\prime$ anomalous dimension.\footnote
{We warn the reader that we have discovered a bug in the implementation of our NLL results.
  The numerical impact is not dramatic and it is discussed in detail in App.~\ref{sec:bug}.
  All numerical results presented here, including the ones with \texttt{HELL 2.0} settings (i.e., using the LL$'$ anomalous dimension),
  have been obtained with the corrected implementation of the resummation.}

\begin{figure}[t]
  \centering
  \includegraphics[width=0.495\textwidth,page=3]{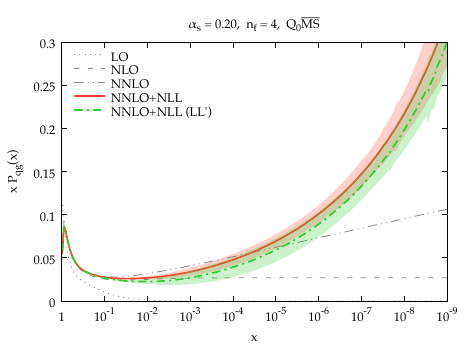}
  \includegraphics[width=0.495\textwidth,page=4]{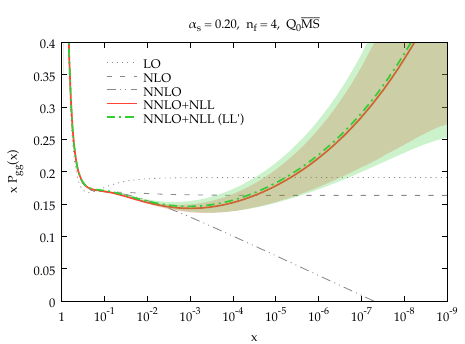}
  \caption{Comparison between resummed and matched splitting functions in two variants of small-$x$ resummation.}
  \label{fig:comp}
\end{figure}
To better appreciate similarities and differences, we compare the two variants of the resummation in Fig.~\ref{fig:comp},
focussing on $P_{qg}$ on the left and on $P_{gg}$ on the right.
The current default, denoted with ``NNLO+NLL'' (solid red) is compared to the choice we made in Ref.~\cite{Bonvini:2017ogt},
which has been labelled ``NNLO+NLL (LL$'$)'' (dot-dashed green). For completeness, we also show fixed-order results (gray).
We see that the difference for $P_{gg}$ are very small and well within the uncertainty band.
The difference for $P_{qg}$ is not large either, even though the uncertainty bands are smaller and comparable in size with such a difference.
Similar considerations hold for the other splitting functions because,
as it is clear from Fig.~\ref{fig:Pres}, $P_{qq}$ and $P_{gq}$ behave similarly to
$P_{qg}$ and $P_{gg}$ at small $x$, due to the colour-charge relations, Eqs.~\eqref{eq:Delta4gamma}.
We have checked that the comparison between the predictions obtained using the NLL anomalous dimension
versus the LL$^\prime$ remains equivalent also at NLO+NLL accuracy.

A striking feature of the result is the small size of the uncertainty band that we obtain for $P_{qg}$.
This is rather counterintuitive because $P_{qg}$ (and $P_{qq}$) start at NLL
and they are therefore known only at their leading non-vanishing logarithmic accuracy.
Perhaps this signals the limitation in our way of estimated theoretical uncertainties,
which is currently purely based on the variation of subleading contributions related to the running of the strong coupling.
Thus, in order to better assess the impact of subleading logarithms,
it would be important to push the accuracy of the resummation,
at least for the quark-initiated splitting functions, one logarithmic order higher. 

\begin{figure}[t]
  \centering
  \includegraphics[width=0.495\textwidth,page=9]{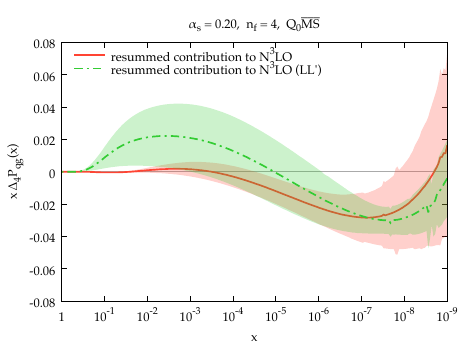}
  \includegraphics[width=0.495\textwidth,page=10]{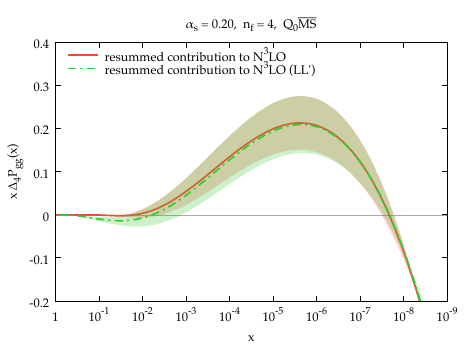}
  \caption{Comparison between the resummed contributions $\Delta_4P_{qg}$ (left) and $\Delta_4P_{gg}$ (right)
    to be added to N$^3$LO splitting functions (when available) in two variants of small-$x$ resummation.}
  \label{fig:Delta4P}
\end{figure}

We now consider the resummation matched to one order higher, in view a future combination with N$^3$LO splitting functions.
Also in this case, we compare the two variants of resummation, which led to very similar results when matched to NNLO.
We have already argued from theoretical grounds that usage of the LL$'$ variant is less favourable
because one has less control over the subleading poles that appear in the expansion of the resummation. 
In Fig.~\ref{fig:Delta4P} we see that this worry is indeed justified.
In these plots we show the resummed contributions
$x \Delta_4 P_{qg}$ (on the left) and $x \Delta_4 P_{gg}$ (on the right), which would have
to be added to the N$^3$LO splitting functions according to Eq.~\eqref{eq:match}.
The solid red curve denotes the default resummation in \texttt{HELL~3.0} based on the NLL anomalous dimension,
while in dot-dashed green we show the LL$'$ variant. 
Both plots show an issue of the LL$'$, which was absent when matching at lower orders:
the resummed contributions give rise to a (most likely) spurious contribution at moderate $x$,
which is instead absent if full NLL is employed in the resummation.
Indeed, for $x\gtrsim 10^{-3}\div10^{-2}$ we expect to be outside the resummation region,
and the effect of resummation should be smaller compared to
the fixed-order contribution, which is more reliable in this region.
This behaviour is violated in $\Delta_4 P_{ij}$ when using the LL$'$ anomalous dimension,
due a large contribution of $\gamma_2^{\rm LL'}$, Eq.~\eqref{eq:gamma2LLp}, entering in Eq.~\eqref{eq:gammaqg},
which makes $\Delta_4 P_{ij}$ even larger than $\Delta_3 P_{ij}$ in this region, despite it being of higher order in $\as$.
Consistently, the green curve also has a rather large uncertainty band in that region,
which makes it almost compatible with the red curve, which has instead a smaller uncertainty, as one would expect.
We interpret this behaviour as further confirmation of the aforementioned theoretical arguments in favour of our new default. 
However, it should be stressed once again that both curves feature the same logarithmic accuracy,
and hence this discrepancy contributes to our theoretical uncertainty.

\subsection{Approximate N$^3$LO}

\begin{figure}[t]
  \centering
  \includegraphics[width=0.495\textwidth,page=11]{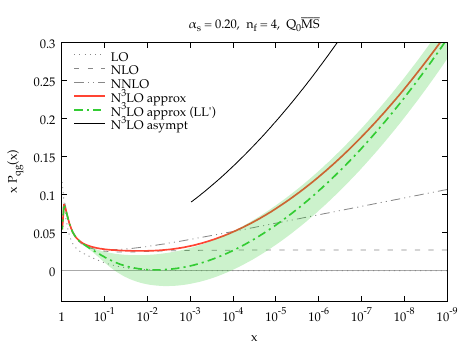}
  \includegraphics[width=0.495\textwidth,page=14]{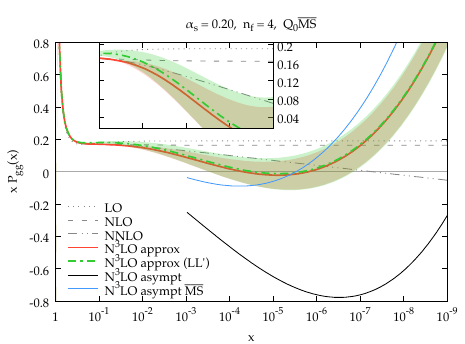}
  \caption{Approximate N$^3$LO prediction (solid) of the $P_{qg}$ (left) and $P_{gg}$ (right) splitting functions
    as obtained by the $\Ord(\as^4)$ expansion of the resummation.
    The NLL asymptotic behaviour is also shown (light solid).}
  \label{fig:PN3LO}
\end{figure}

We can use the expansion of the resummed splitting functions to $\Ord(\as^4)$
to make an approximate prediction of the N$^3$LO splitting functions,
simply by adding it to the exact NNLO ones.
This is shown in solid red in Fig.~\ref{fig:PN3LO} for $P_{qg}$ (left plot) and $P_{gg}$ (right plot),
and in dot-dashed green for the LL$'$ variant.
According to the discussion in the previous section, the latter curve is not expected to be accurate
in the region of moderately large $x$, where it has an unphysically large effect.
These predictions are further supplemented by the same uncertainty band that appears on resummed results,
which happens to be invisible for $P_{qg}$ when using our new default implementation\footnote
{Indeed, at this order, our uncertainty band originates from the parameter $T$ appearing in Eq.~\eqref{eq:gammaqg},
  and on the potential dependence of the anomalous dimensions $\gamma_{0,1,2}$ on the parameter $T'$ defined in App.~\ref{sec:expansion}.
  Since the NLL anomalous dimension is more precise than the LL$'$ one, none of the $\gamma^{\rm NLL}_{0,1,2}$ depends on $T'$,
  while $\gamma^{\rm LL'}_2$ does (and also $\gamma^{\rm NLL}_3$, which contributes to $P_{gg}$).
  It is the latter ($T'$) dependence that generates the uncertainty bands in Fig.~\ref{fig:PN3LO},
  while the $T$ parameter variation has no appreciable effect at this order on any of the two variants.}
(and thus confirms that our uncertainty band underestimates the actual uncertainty on $P_{qg}$).
Additionally, we also show in light solid black the asymptotic small-$x$ behaviour at N$^3$LO,
as obtained by adding to the exact NNLO the pure NLL contributions, Eq.~\eqref{eq:gamma3sx}, without any subleading effects.
Note that these results are obtained in the $Q_0\MSbar$ scheme.  
While we have not implemented the resummation (and hence its expansion) in $\MSbar$, we can easily plot the asymptotic behaviour
of the splitting functions in this scheme, exploiting the results of Sect.~\ref{sec:MSbar}.
The quark splitting function $P_{qg}$ is unaffected, while the $\MSbar$ asymptotic result for $P_{gg}$,
Eq.~\eqref{eq:gamma3sxMSbar}, is shown in solid cyan.

In the small-$x$ limit, approximate predictions behave as their asymptotic expansions, by construction.
The difference is due to subleading NNLL contributions, behaving as $\frac1x\log\frac1x$ at this order
(a straight line in the plots).
While these NNLL contributions are subleading at asymptotically small $x$, their effect is sizeable
for all the $x$ range shown in the plots, which is rather large, reaching $x=10^{-9}$.
This is true in particular for $P_{gg}$, where the pure NLL asymptotic curve is very different from
the approximate N$^3$LO, so much that in order to display the asymptotic behaviour we had to plot
$xP_{gg}(x)$ in a rather extended range. For this reason, we also added an inset which zooms in to the region $10^{-5}<x<0.1$,
most relevant for HERA and LHC physics, to better appreciate the perturbative behaviour of the fixed-order splitting functions.
This shows once again that subleading contributions have a very important role
at intermediate and moderately small values of $x$. 
Similar conclusions were reached some time ago in Ref.~\cite{Blumlein:1997em}
(see also Ref.~\cite{Blumlein:1995jp} for a similar study in the non-singlet case).
This also suggests that the approximate N$^3$LO prediction
that we plotted has a huge uncertainty, likely larger than what we estimate with our uncertainty bands.

By comparing resummed results (Fig.~\ref{fig:Pres}) and their expansions (Fig.~\ref{fig:comp}),
we can conclude that, while the small-$x$ contributions to $P_{qg}$, as obtained from the expansion of the resummation,
behave in a perturbative way, the prediction of the N$^3$LO contribution to $P_{gg}$,
which is more directly sensitive to the (perturbatively unstable) BFKL kernel,
is very different from its all-order counterpart, as it was the NNLO contribution.
We must conclude that these approximate N$^3$LO predictions cannot be regarded as a faithful estimate
of the actual N$^3$LO (especially for $P_{gg}$ and $P_{gq}$) due to potentially underestimated subleading contributions.
However, what we can certainly conclude is that exact N$^3$LO evolution (when available)
will be unreliable at small $x$, and thus it will necessarily have to be supplemented
with the all-order resummation of small-$x$ contributions.

As we have argued before, because of the sensitivity of the N$^3$LO splitting functions
to subleading logarithmic contributions, it would be important to push the resummation to NNLL accuracy.
However, NNLL resummation requires at least the knowledge of the NNLO BFKL kernel,
which is so far only known in a collinear approximation~\cite{Marzani:2007gk}.
It will be important in the future to explore the possibility of computing the BFKL kernel to NNLO~\cite{DelDuca:2008jg,Bret:2011xm,DelDuca:2011ae,DelDuca:2014cya,Caron-Huot:2016tzz,Caron-Huot:2017fxr},
and perhaps to consider the option of using its collinear approximation.

Finally, it interesting to note that the asymptotic behaviour in $\MSbar$ appears to be closer to the all-order result (albeit computed in a different scheme). 
This suggests that a future study of the resummation in $\MSbar$ may reveal interesting properties in terms of the size of subleading contributions, despite the fact that in this scheme we expect stronger cancellations between coefficient functions and parton evolution.

\section{Conclusions}
\label{sec:conclusions}
In this paper we have discussed the role of higher-order corrections to the splitting functions,
which govern the evolution of the parton distribution functions. 
In particular, we have exploited results in small-$x$ resummation to study the behaviour
of the yet-unknown four-loop splitting functions in the singlet sector.
Our results stress once again the fact that small-$x$ singularities lead to loss of perturbative stability,
when higher orders are considered.
This has been masked so far by accidental cancellations at NLO
but it becomes apparent at NNLO, even though also there the strongest singularity is accidentally zero,
thus slightly mitigating the perturbative deterioration.
Instead, at the next orders there are no accidental cancellations,
so the perturbative instability is no longer moderated,
and we estimate it to be rather severe at N$^3$LO.
Thanks to this work, this potential instability can be solved by adding
the resummation to the full four-loop splitting functions, when these will become available.

We have also investigated the possibility of using the expansion of the resummation
to construct approximate N$^3$LO splitting functions.
Unfortunately, we have found that subleading corrections, which are only partially included in our approach,
have a sizeable impact at moderate $x$, thus rendering the construction of approximate fixed-order
splitting functions rather uncertain.
However, our asymptotic results can be used as a check on the full four-loop calculation,
or for complementing an approximate computation based on integer Mellin moments.

While performing these studies we have encountered a potential source of instability
in the way the resummation of the quark anomalous dimension $\gamma_{qg}$ was implemented
in \texttt{HELL 2.0} in Ref.~\cite{Bonvini:2017ogt}, which was based on a hybrid resummation formula denoted LL$^\prime$.
Therefore, we have adopted as a new default a resummation fully based on NLL and consequently
released a new version of the resummation code \texttt{HELL 3.0}:
\begin{center}
  \href{http://www.ge.infn.it/~bonvini/hell}{\tt www.ge.infn.it/$\sim$bonvini/hell}
\end{center}
As the distinction between the two choices is beyond the accuracy of the calculation,
the old option can be, and should be, still used to estimate theoretical uncertainties.
Furthermore, we anticipate that an analogous situation appears in the resummation of partonic coefficient functions.
This issue will be discussed in a forthcoming study~\cite{Bonvini:2018iwt}.
Finally, these results have been recently exploited in a double-resummed calculation
of the Higgs production cross section~\cite{Bonvini:2018ixe}.

\acknowledgments
{
We thank Andreas Vogt for encouraging us to write up these results and Richard Ball for useful discussions.
The work of MB is supported by the Marie Sk\l{}odowska Curie grant HiPPiE@LHC.
}

\appendix
\section{\boldmath Perturbative expansion of resummed anomalous dimensions to N$^3$LO}
\label{sec:expansion}

In this appendix we derive the expansion in powers of $\as$ of the resummed plus eigenvalue
of the singlet anomalous dimension matrix presented in Ref.~\cite{Bonvini:2017ogt}.
Specifically, we provide the detailed computation of the expansion of the NLL resummed anomalous dimensions up to $\Ord(\as^4)$,
and of the LL$^\prime$ anomalous dimension up to $\Ord(\as^3)$, as needed for matching NLL resummation in DGLAP evolution to N$^3$LO.
We recall that the anomalous dimensions at LL$^\prime$ and NLL are constructed as~\cite{Bonvini:2017ogt}
\begin{subequations}\label{eq:gammaresNLLdef}
\begin{align}
\gamma_+^{\rm res\,LL'}(N,\as)
&= \gamma_\text{DL-LO} (N,\as) + \Delta_\text{DL-LO}\gamma_{\rm rc}^{\rm NLL}(N,\as) + \gamma_{\rm match}^{\rm LO+LL'} (N,\as) \nonumber\\
&\quad - \[\Delta_\text{DL-LO}\gamma_{\rm rc}^{\rm NLL}(1,\as) + \gamma_{\rm match}^{\rm LO+LL'} (1,\as) \] f_{\rm mom}(N), \\
\gamma_+^{\rm res\,NLL}(N,\as)
&= \gamma_\text{DL-NLO}(N,\as) + \Delta_\text{DL-NLO}\gamma_{\rm rc}^{\rm NLL}(N,\as) + \Delta\gamma_{ss}^{\rm rc}(N,\as) + \gamma^{ss}_{\rm match}(N,\as) \nonumber\\
&\quad - \[\Delta_\text{DL-LO}\gamma_{\rm rc}^{\rm NLL}(1,\as) + \Delta\gamma_{ss}^{\rm rc}(1,\as) + \gamma^{ss}_{\rm match}(1,\as)\] f_{\rm mom}(N),
\end{align}
\end{subequations}
where $\gamma_\text{DL-(N)LO}$ is the double-leading (DL) anomalous dimension,
$\Delta_\text{DL-(N)LO}\gamma_{\rm rc}^{\rm NLL}$ is the contribution (to be added to the DL) coming from the resummation of running-coupling (RC) effects,
$\Delta\gamma_{ss}^{\rm rc}$ is a running-coupling correction to the fixed-coupling DL construction at NLL,
$\gamma_{\rm match}^{\rm LO+LL'}$ and $\gamma^{ss}_{\rm match}$ are matching functions to cancel mismatched singularities,
and the second line of each equation restores momentum conservation, i.e.\ the constraint $\gamma_+^{\rm res\,LL'}(1,\as)=\gamma_+^{\rm res\,NLL}(1,\as)=0$,
through a subleading function $f_{\rm mom}$ defined in Eq.~\eqref{eq:fmom}.
All these ingredients have been presented in Ref.~\cite{Bonvini:2017ogt} and will be used in the following.

Before starting, we recall that the DL anomalous dimension is constructed starting from the fixed-order
BFKL kernel matched to the fixed-order anomalous dimension. Thus, one of the ingredients of (N)LL
resummation is the (N)LO anomalous dimension.
In Ref.~\cite{Bonvini:2017ogt} an approximate form for the input anomalous dimension was suggested
to facilitate the numerical implementation and to solve a potential issue. The approximation
does not represent any loss of accuracy, as the only requirements needed for the input anomalous dimension are to
be accurate at NLL and to conserve momentum, both of which are satisfied in the approximation of Ref.~\cite{Bonvini:2017ogt}.
At LO and NLO they are given by
\begin{subequations}\label{eq:gammaapprox}
\begin{align}
\gammaa_0 &= \frac{a_{11}}{N} + a_{10} -\frac{2 (a_{11}+a_{10}) N}{N+1}, \\
\gammaa_1 &= \frac{a_{21}}{N} + a_{20} -\frac{2 (a_{21}+a_{20}) N}{N+1},
\end{align}
\end{subequations}
with
\begin{subequations}\label{eq:aij}
\begin{align}
  a_{11} &= \frac{C_A}\pi, \\
  a_{21} &= n_f\frac{26C_F-23C_A}{36\pi^2}, \\
  a_{10} &= -\frac{11C_A + 2n_f(1-2C_F/C_A)}{12\pi}, \\
  a_{20} &= \frac1{\pi^2}\[\frac{1643}{24} - \frac{33}2 \zeta_2 - 18 \zeta_3 + n_f\(\frac49\zeta_2-\frac{68}{81}\) + n_f^2\frac{13}{2187}\].
\end{align}
\end{subequations}
In the next, we start from LL$^\prime$ resummation, and then move to NLL.
The computation follows closely the one presented in Sect.~3 and 4 of Ref.~\cite{Bonvini:2017ogt},
extending it to one extra order.

\subsection{Expansion of the LL$^\prime$ anomalous dimension}

We start expanding the LL$^\prime$ anomalous dimension up to $\Ord(\as^3)$.
The first ingredient for resummation is the DL resummed anomalous dimension $\gamma_{\rm DL}$,
which is obtained from the implicit equation
\beq\label{eq:duality}
\chi_{\Sigma}\(\gamma_{\rm DL}(N,\as),N,\as\) = N.
\eeq
The function $\chi_{\Sigma}\(M,N,\as\)$ is the so-called off-shell BFKL kernel~\cite{Altarelli:2005ni,Bonvini:2016wki}.
For LL resummation it is given by~\cite{Altarelli:2005ni}
\beq\label{eq:chiDLLO}
\chi_{\Sigma}^{\rm LO}(M,N,\as)
= \chi_{s}\(\frac{\as}M\) + \chi_{s}\(\frac{\as}{1-M+N}\) + \as \tilde\chi_0(M,N) + c_{\rm mom}^{\rm LO}(\as) f_{\rm mom}(N),
\eeq
where the function $\chi_s(\as/M)$ is the dual of the LO anomalous dimension $\as\gammaa_0$,
\beq\label{eq:chis}
\as\gammaa_0\(\chi_s\(\frac{\as}{M}\)\) = M \qquad \Leftrightarrow\qquad \chi_s\(\frac1{\gammaa_0(N)}\)=N,
\eeq
and
\beq\label{eq:chi0tilde}
\tilde\chi_0(M,N) = \chi_{01} \Big[\psi(1) + \psi(1+N) - \psi(1+M) - \psi(2-M+N)\Big]
\eeq
is the off-shell extension of the LO BFKL kernel after subtracting double counting with $\chi_s$.
The last term restores the momentum conservation constraint $\gamma_{\rm DL}(1,\as)=0$, namely by duality $\chi_\Sigma(0,1,\as)=1$,
through a function
\beq\label{eq:fmom}
f_{\rm mom}(N) = \frac{4N}{(N+1)^2},
\eeq
and with the coefficient
\beq
c_{\rm mom}^{\rm LO}(\as) = -\chi_{s}\(\frac{\as}{2}\) - \as \tilde\chi_0(0,1).
\eeq
The coefficient $\chi_{01}$ appearing in Eq.~\eqref{eq:chi0tilde} is the first of the expansion of $\chi_s$,
\beq\label{eq:chisser}
\chi_s\(\frac{\as}{M}\) = \sum_{k=1}^\infty \chi_{0k} \(\frac{\as}M\)^k.
\eeq
All $\chi_{0k}$ coefficients are determined in terms of $a_{11}$ and $a_{10}$, Eq.~\eqref{eq:aij},
through Eq.~\eqref{eq:chis} and Eq.~\eqref{eq:gammaapprox}. In particular, the first three coefficients are given by
\begin{align}\label{eq:chi0k}
\chi_{01} &= a_{11}, &
\chi_{02} &= a_{11}a_{10}, &
\chi_{03} &= a_{11}(a_{10}^2-2a_{11}a_{10}-2a_{11}^2).
\end{align}
Following Ref.~\cite{Bonvini:2017ogt}, we write the $\as$-expansion of the DL anomalous dimension
\beq\label{eq:gammaDLLOexp}
\gamma_\text{DL-LO}(N,\as) = \as\gammaa_0(N) +\as^2\gammat_1(N) +\as^3\gammat_2(N)+ \Ord(\as^4),
\eeq
where $\gammaa_0$ is the input LO anomalous dimension Eq.~\eqref{eq:gammaapprox}
used in the definition of $\chi_s$, Eq.~\eqref{eq:chis},
while $\gammat_1$ and $\gammat_2$ are the predictions of the resummation that we aim to find.
Then, we substitute it into Eq.~\eqref{eq:duality} with $\chi_\Sigma$ given in Eq.~\eqref{eq:chiDLLO}, and expand the equation in powers of $\as$.
The most delicate function to expand is the collinear $\chi_s$ in Eq.~\eqref{eq:chiDLLO}, for which we find (omitting arguments to facilitate reading)
\begin{align}\label{eq:chisexpansion}
\chi_s\(\frac{\as}{\gamma_\text{DL-LO}}\)
&= \chi_s\(\frac1{\gammaa_0}\[1-\as\frac{\gammat_1}{\gammaa_0}+\as^2\frac{\gammat_1^2-\gammaa_0\gammat_2}{\gammaa_0^2}+\Ord(\as^3)\]\) \nonumber\\
&= \chi_s\(\frac{1}{\gammaa_0}\) - \as\frac{\gammat_1}{\gammaa_0^2}\chi_s'\(\frac{1}{\gammaa_0}\)
  + \as^2\[\frac{\gammat_1^2-\gammaa_0\gammat_2}{\gammaa_0^3}\chi_s'\(\frac{1}{\gammaa_0}\)
  +\frac{\gammat_1^2}{2\gammaa_0^4}\chi_s''\(\frac{1}{\gammaa_0}\)\]
  + \Ord(\as^3)\nonumber\\
&= N + \as\frac{\gammat_1}{\gammaa_0'}
  + \as^2\[\frac{\gammat_2}{\gammaa_0'}-\frac{\gammat_1^2\gammaa_0''}{2\gammaa_0'^3}\] + \Ord(\as^3),
\end{align}
where in the last equality we have used the definition Eq.~\eqref{eq:chis}, and the formulae for the derivatives
\beq
\chi_s'\(\frac{1}{\gammaa_0}\) = -\frac{\gammaa_0^2}{\gammaa_0'},\qquad
\chi_s''\(\frac{1}{\gammaa_0}\) = \frac{\gammaa_0^2}{\gammaa_0'}\[2\gammaa_0-\frac{\gammaa_0^2\gammaa_0''}{\gammaa_0'^2}\],
\eeq
which can be derived from the very same definition.
The prime $\prime$ denotes a derivative with respect to the argument of the function, so $\chi_s'(1/\gammaa_0)$ is a derivative with respect to $1/\gammaa_0$,
and $\gammaa_0''$ is a double derivative with respect to $N$.
The anticollinear $\chi_s$ gives instead
\beq
\chi_s \(\frac{\as}{1-\gamma_\text{DL-LO}+N}\)
= \as\frac{\chi_{01}}{1+N} + \as^2 \frac{\chi_{02}+\chi_{01}\gammaa_0}{(1+N)^2} + \Ord(\as^3).
\eeq
The kernel Eq.~\eqref{eq:chiDLLO} expands as
\beq
\tilde\chi_0(\gamma_\text{DL-LO}(N,\as),N) = \tilde\chi_0(0,N) + \as \gammaa_0 \tilde\chi_0'(0,N) + \Ord(\as^2),
\eeq
where the derivative is with respect to $M$, i.e.\ the first argument.
Putting everything together Eq.~\eqref{eq:duality} brings to the expanded equality
\begin{align}
N &= N +\as\[\frac{\gammat_1}{\gammaa_0'} + \frac{\chi_{01}}{1+N} + \tilde\chi_0(0,N) - \(\frac{\chi_{01}}{2} + \tilde\chi_0(0,1)\)f_{\rm mom}(N)\] \nonumber\\
&\qquad\;+ \as^2 \[
\frac{\gammat_2}{\gammaa_0'}-\frac{\gammat_1^2\gammaa_0''}{2\gammaa_0'^3}
+\frac{\chi_{02}+\chi_{01}\gammaa_0}{(1+N)^2}
+\gammaa_0 \tilde\chi_0'(0,N)
-\frac{\chi_{02}}{4} f_{\rm mom}(N)
\]\nonumber\\
&\qquad\;+ \Ord(\as^3),
\end{align}
from which it immediately follows
\begin{align}
\gammat_1(N) &= -\gammaa_0'(N) \[ \frac{\chi_{01}}{1+N} + \tilde\chi_0(0,N)
 -\(\frac{\chi_{01}}2 + \tilde\chi_0(0,1)\)f_{\rm mom}(N) \],\\
\gammat_2(N) &= \frac{\gammat_1^2(N)\gammaa_0''(N)}{2\gammaa_0'^2(N)}
-\gammaa_0'(N) \[ \frac{\chi_{02}+\chi_{01}\gammaa_0(N)}{(1+N)^2}
+\gammaa_0(N) \tilde\chi_0'(0,N)
-\frac{\chi_{02}}{4} f_{\rm mom}(N)\].
\end{align}
Note that the $\Ord(\as^0)$ term cancels automatically, because $\gammaa_0$ in Eq.~\eqref{eq:gammaDLLOexp} is the one used in the definition of $\chi_s$,
Eq.~\eqref{eq:chis}.
The expansion terms of the off-shell kernel $\tilde\chi_0(M,N)$, Eq.~\eqref{eq:chi0tilde}, are given by
\beq\label{eq:chi00}
\tilde\chi_0(0,N) = -\frac{\chi_{01}}{1+N}, \qquad
\tilde\chi_0'(0,N) = -\frac{\chi_{01}}{(1+N)^2} + \chi_{01}\[\psi_1(1+N)-\zeta_2\],
\eeq
which lead to the predictions
\begin{align}
\gammat_1(N) &= 0, \\
\gammat_2(N) &= -\gammaa_0'(N)
  \[ \frac{\chi_{02}}{(1+N)^2}+\chi_{01}\[\psi_1(1+N)-\zeta_2\]\gammaa_0(N)-\frac{\chi_{02}}{4} f_{\rm mom}(N)\].
\end{align}
As already noted in Ref.~\cite{Bonvini:2017ogt}, the fact that $\gammat_1$ vanishes is not surprising:
indeed, the LL pole of the exact NLO $\gamma_+^{(1)}$ and NNLO $\gamma_+^{(2)}$ are accidentally zero,
so the only part which is supposed to be predicted correctly by LL resummation was indeed expected to vanish.

Having computed the expansion of the DL part, we now move to the RC contributions.
The function that resums the running-coupling effects is given by~\cite{Bonvini:2017ogt}
\beq\label{eq:RCsol}
\gamma_{\rm rc}(N,\as) = M_{\rm min} + \beta_0\asb \[ z \frac{k_{\nu}'(z)}{k_{\nu} (z)} -1 \],
\eeq
where $k_\nu(z)$ is a Bateman function, with
\begin{subequations}\label{eq:BatemanFunctions}
\begin{align}
  \frac1{\asb} &= \frac1{\as} + \frac{\kappa'-2c'/M_{\rm min}^2}{\bar\kappa+2(N-\bar c)/M_{\rm min}^2}\\
  z &= \frac1{\beta_0\asb}\sqrt{\frac{N-\bar c}{\bar\kappa/2 +(N-\bar c)/M_{\rm min}^2}}\\
  \nu &= \(\frac{c'}{N-\bar c} + \frac{\kappa'-2c'/M_{\rm min}^2}{\bar\kappa+2(N-\bar c)/M_{\rm min}^2}\) \asb z, \\
  \bar c(\as) &= c(\as) - \as c'(\as), \qquad 
  \bar \kappa(\as) = \kappa(\as) - \as \kappa'(\as).
\end{align}
\end{subequations}
In the equations above, $M_{\rm min}(\as)$ is the position of the minimum of the BFKL kernel,\footnote
{Note that in Ref.~\cite{Bonvini:2017ogt} we suggested to compute the $M_{\rm min}$ from the kernel
in symmetric variables, while we now decided to use the one in DIS variables,
as we discuss in greater detail in Sect.~\ref{sec:coefficients}.}
and $c(\as)$ and $\kappa(\as)$ are the value and curvature of the kernel at such minimum.
In deriving Eq.~\eqref{eq:RCsol}, the $\as$-dependence of the BFKL kernel has been approximated linearly,
keeping the value of the kernel and its $\as$-derivative correct.
In Ref.~\cite{Bonvini:2017ogt} a variant of this approximation in which the kernel is assumed to be proportional to $\as$
(i.e., as if it was a purely LO kernel) was considered to study the impact of subleading logarithmic contributions.
This variant is recovered by letting $c'\to c/\as$, $\kappa'\to\kappa/\as$, and represents an equally valid alternative.
The RC contribution to be added to the DL-LO result is given by
\begin{align}\label{eq:DeltaRCLO}
\Delta_\text{DL-LO}\gamma_{\rm rc}(N,\as)
&\equiv \gamma_{\rm rc}(N,\as) - \[M_{\rm min} - \sqrt{\frac{N-c}{\kappa/2 +(N-c)/M_{\rm min}^2}} - \beta_0\as\]
\nonumber\\
&= \beta_0\as^2 \frac{3\kappa_0/32-c_0}N\nonumber\\
&\quad+ \beta_0\as^3 \[c_0\frac{3\kappa_0/32-c_0}{N^2}
  + T'\frac{3\kappa_1/32 -c_1}{N}
  + \kappa_0\frac{\beta_0 + 6 m_1}{16N}
  + \kappa_0\frac{24 c_0 - 3 \kappa_0}{256N^2}
\]\nonumber\\
&\quad+ \Ord(\as^4),
\end{align}
where $\kappa_{0,1}$ and $c_{0,1}$ are the $\Ord(\as^{1,2})$ terms of $\kappa$ and $c$,
and $m_1$ is determined from $M_{\rm min}(\as)=1/2 + \as m_1+\Ord(\as^2)$.
To cover both $\as$-dependence approximations, we have introduced a parameter $T'$
which equals $2$ in the default approximation and equals $1$ in the limit $c'\to c/\as$, $\kappa'\to\kappa/\as$.
The values of $\kappa_1$, $c_1$ and $m_1$ depend on the actual kernel used.
For LL$^\prime$ resummation, the DL-NLO is used for RC resummation, differently from pure LL resummation
which uses the DL-LO kernel.
Thus, the second order coefficients in Eq.~\eqref{eq:DeltaRCLO} are $m_1^{\rm NLO}$, $c_1^{\rm NLO}$ and $\kappa_1^{\rm NLO}$,
explicitly given in Sect.~\ref{sec:coefficients}.
Because in LL$^\prime$ resummation the RC contributions, computed from the NLO BFKL kernel, are matched to the DL-LO anomalous dimension,
there is a mismatch in the singularities at small $x$ which are cured by the matching function
$\gamma_{\rm match}^{\rm LO+LL'}$ defined in Ref.~\cite{Bonvini:2017ogt}.
Its expansion gives
\beq
\gamma_{\rm match}^{\rm LO+LL'} (N,\as)
= \as^3\frac{16c_0\(\delta\kappa_1 + 6\kappa_0\delta m_1\) +\kappa_0\(16\delta c_1-3\delta\kappa_1-15\kappa_0 \delta m_1\)}{512N^2}+\Ord(\as^4),
\eeq
where
\begin{align}
\delta m_1 &= m_1^{\rm NLO}-m_1^{\rm LO}, \\
\delta c_1 &= c_1^{\rm NLO}-c_1^{\rm LO}, \\
\delta\kappa_1 &= \kappa_1^{\rm NLO} -\kappa_1^{\rm LO},
\end{align}
are the differences between the coefficients computed with the DL-NLO and the DL-LO BFKL kernels.
Explicit results are given in Sect.~\ref{sec:coefficients}.

Putting everything together, the expansion in powers of $\as$ of the full LL$^\prime$ anomalous dimension is given by
\begin{align}
\gamma^{\rm res\, LL'}(N,\as)
&= \as \gammaa_0(N)
+ \as^2\beta_0\(\frac{3}{32}\kappa_0-c_0\)\(\frac1N-f_{\rm mom}(N)\) \nonumber\\
&+ \as^3 \bigg\{ \frac{\lambda_2}{N^2}+\frac{\lambda_1}{N} - \(\lambda_2+\lambda_1\)f_{\rm mom}(N) \nonumber\\
&\qquad\qquad-\gammaa_0'(N) \[ \frac{\chi_{02}}{(1+N)^2}+\chi_{01}\[\psi_1(1+N)-\zeta_2\]\gammaa_0(N)-\frac{\chi_{02}}{4} f_{\rm mom}(N)\]\bigg\}\nonumber\\
&+\Ord(\as^4),
\end{align}
with
\begin{subequations}\label{eq:rhos}
\begin{align}
\lambda_1 &= \beta_0\[\frac{3}{32}T'\kappa_1^{\rm NLO} -T' c_1^{\rm NLO} + \frac{\beta_0 + 6 m_1^{\rm NLO}}{16}\kappa_0\] \label{eq:rho1} \\
\lambda_2 &= \beta_0 \frac{48 c_0 \kappa_0 - 3 \kappa_0^2-256c_0^2}{256}
+\frac{16c_0\(\delta\kappa_1 + 6\kappa_0\delta m_1\) +\kappa_0\(16\delta c_1-3\delta\kappa_1-15\kappa_0 \delta m_1\)}{512}.
\end{align}
\end{subequations}
Using Eq.~\eqref{eq:chi0k} and replacing the explicit form of $\gammaa_0(N)$, Eq.~\eqref{eq:gammaapprox},
we obtain the results presented in Eq.~\eqref{eq:gamma1LLp} and Eq.~\eqref{eq:gamma2LLp}.

\subsection{Expansion of the NLL anomalous dimensions}
\label{sec:DLNLL}

We now move to the NLL anomalous dimension. This time, we need to push our expansion to one order higher.
The off-shell kernel needed for NLL resummation is
\begin{align}\label{eq:chiDLNLO}
\chi_\Sigma^{\rm NLO}(M,N,\as)
  &= \chi_{s,\rm NLO}(M,\as) + \chi_{s,\rm NLO}(1-M+N,\as) \nonumber\\
  &+ \as \tilde\chi_0(M,N)  + \as^2 \tilde\chi_1(M,N) + \as^2\chi_{1}^{\rm corr}(M,N,\as) \nonumber\\
  &+ c_{\rm mom}^{\rm NLO}(\as)f_{\rm mom}(N).
\end{align}
The function $\chi_{s,\rm NLO}(M,\as)$ is the generalization of $\chi_s$ to the next order,
which is obtained as the exact dual of the NLO anomalous dimension,
\beq
\chi_{s,\rm NLO}\big(\as\gammaa_0(N)+\as^2\gammaa_1(N),\as\big) = N.
\eeq
This kernel can be expanded as
\beq\label{eq:chisNLO}
\chi_{s,\rm NLO}(M,\as) = \sum_{j=0}^\infty \as^j \sum_{k=1}^\infty \chi_{jk} \(\frac{\as}M\)^k,
\eeq
which generalizes Eq.~\eqref{eq:chisser}, which is just the $j=0$ part of this function.
All these coefficients are given in terms of $a_{11}$, $a_{10}$, $a_{21}$ and $a_{20}$;
the relevant ones for what follows are
\begin{align}\label{eq:chijk}
\chi_{11} &= a_{21}, &
\chi_{12} &= a_{21}a_{10}+a_{11}a_{20}, &
\chi_{21} &= 0.
\end{align}
The kernel $\tilde\chi_1(M,N)$ was given in Eqs.~(A.23)--(A.29) of Ref.~\cite{Bonvini:2016wki},
and can be written as
\beq
\tilde\chi_1(M,N) = \tilde\chi_1^{\rm u}(M,N) - \tilde\chi_1^{\rm u}(0,N) + \tilde\chi_1^{\rm u}(0,0),
\eeq
where the function
\beq
\tilde\chi_1^{\rm u}(M,N) = \breve\chi_1(M,N)
- \chi_{02}\(\frac{1}{M^2}+\frac{1}{(1-M+N)^2}\)
- \chi_{11}\(\frac{1}{M}+\frac{1}{1-M+N}\)
\eeq
is regular in $M=0$ and $1-M+N=0$. The function $\breve\chi_1$ is given by~\cite{Bonvini:2016wki}
\begin{align} \label{eq:chi1breve}
\breve\chi_1(M,N)
&= -\frac{1}{2} \chi_0(M,N)\, \chi_{01}\Big[2\psi_1(1+N)-\psi_1(M) -\psi_1(1-M+N)\Big]\nonumber\\
&\quad -\frac{1}{2}\beta_0\chi_{01}\[\(\frac{\chi_0(M,N)}{\chi_{01}}\)^2-\psi_1(M)-\psi_1(1-M+N)\] \nonumber \\
&\quad +\frac{\chi_{01}^2}{4}\Bigg\{\(\frac{12\chi_{11}-26\chi_{02}}{3\chi_{01}^2}-\frac12-2\zeta_2\)
  \Big[\psi (1)-\psi(M)\Big] \nonumber\\
&\qquad\qquad+3 \zeta (3) +\psi_2(M)
+ \zeta_2\[\psi\(\tfrac{1+M}{2}\)-\psi\(\tfrac{M}{2}\)\]+4\phi_L^+(M)\nonumber \\
&\qquad\qquad+\frac{3}{4(1-2M)}\[\psi_1(\tfrac{1+M}{2})-\psi_1(\tfrac{M}{2})+\psi_1(\tfrac{1}{4})-\psi_1(\tfrac{3}{4})\] \nonumber \\
&\qquad\qquad-\(\frac{9}{2}+\frac{6\chi_{02}}{\chi_{01}^2}\)\frac{2+3M(1-M)}{16}
\Bigg[ \frac{\psi_1(\tfrac{1+M}{2})-\psi_1(\tfrac{M}{2})+\psi_1(\tfrac{1}{4})-\psi_1(\tfrac{3}{4})}{1-2M} \nonumber \\
&\hspace{18em} +\frac{\psi_1(\tfrac{1+M}{2})-\psi_1(\tfrac{M}{2})+\psi_1(-\tfrac{1}{4})-\psi_1(\tfrac{1}{4})}{2(1+2M)} \nonumber \\
&\hspace{18em} -\frac{\psi_1(\tfrac{1+M}{2})-\psi_1(\tfrac{M}{2})+\psi_1(\tfrac{3}{4})-\psi_1(\tfrac{5}{4})}{2(3-2M)}\Bigg] \nonumber \\
&\qquad\qquad + (M\leftrightarrow 1-M+N) \Bigg\},
\end{align}
with
\begin{align}
  \chi_0(M,N) &= \chi_{01}\Big[\psi(1) + \psi(1+N) - \psi(M) - \psi(1-M+N)\Big], \label{eq:chi0} \\
  \phi_{L}^{+}(M) &= \int_{0}^{1}d x\;  x^{M-1}\, \frac{\text{Li}_2(x)}{1+x}.\label{eq:phiL}
\end{align}
The extra term $\chi_{1}^{\rm corr}(M,N)$ was corrected in Ref.~\cite{Bonvini:2017ogt}; however,
there was another issue, which we have discovered only now. This issue has an effect at NLL,
namely the claimed accuracy, but it manifests itself (at NLL) only in terms of $\Ord(\as^4)$ and beyond of the anomalous dimension
(for this reason, the comparison of the expansion of the anomalous dimension with the exact one at three loops was successful).
The origin of the issue and our solution are discussed in detail in Sect.~\ref{sec:bug}. The actual expression that we use here
is given by
\begin{align}\label{eq:chi1corr}
\chi_1^{\rm corr}(M,N,\as) &= \beta_0\bigg[-\chi_{01}\psi_1(1-M+N) + \frac{\chi_0(M,N)}M + \chi_0'(M,N) \nonumber\\
&\qquad\qquad-\frac{(1-M+N)^{-1}}{1+N}\(\chi_{01}-\chi'_s\(\frac{\as}{1-M+N}\)\)\bigg].
\end{align}
Finally, the momentum conservation coefficient is given by
\beq
c_{\rm mom}^{\rm NLO} = -\chi_{s,\rm NLO}(2,\as) - \as \tilde\chi_0(0,1) - \as^2 \tilde\chi_1(0,1) - \as^2\chi_{1}^{\rm corr}(0,1,\as).
\eeq
We now consider the expansion of the DL-NLO anomalous dimension
\beq\label{eq:gammaDLNLOexp}
\gamma_\text{DL-NLO}(N,\as) = \as\gammaa_0(N)+\as^2\gammaa_1(N) +\as^3\gammat_2(N) +\as^4\gammat_3(N) + \Ord(\as^5),
\eeq
where both $\gammaa_0(N)$ and $\gammaa_1(N)$ are now an input, Eq.~\eqref{eq:gammaapprox},
and $\gammat_2(N)$ and $\gammat_3(N)$ are the objects that we aim to compute.
The expansion of the collinear $\chi_{s,\rm NLO}$ proceeds as in Eq.~\eqref{eq:chisexpansion}, and gives
\beq\label{eq:chisNLOexpa}
\chi_{s,\rm NLO}(\gamma_\text{DL-NLO},\as) = N + \as^2\frac{\gammat_2}{\gammaa_0'}
+\as^3\[\frac{\gammat_3}{\gammaa_0'}-\frac{\gammat_2\gammaa_1'}{\gammaa_0'^2}\] + \Ord(\as^4).
\eeq
The anticollinear $\chi_{s,\rm NLO}$ expands as
\begin{align}
\chi_{s,\rm NLO} \(1-\gamma_\text{DL-NLO}+N,\as\)
&= \as\frac{\chi_{01}}{1+N} + \as^2 \[\frac{\chi_{02}+\chi_{01}\gammaa_0}{(1+N)^2} + \frac{\chi_{11}}{1+N}\]\nonumber\\
&+ \as^3\[\frac{\chi_{03}+2\chi_{02}\gammaa_0+\chi_{01}\gammaa_0^2}{(1+N)^3}
  + \frac{\chi_{12}+\chi_{11}\gammaa_0+\chi_{01}\gammaa_1}{(1+N)^2} + \frac{\chi_{21}}{1+N}\]\nonumber\\
&+ \Ord(\as^4).
\end{align}
The kernels give instead
\begin{align}
\tilde\chi_0(\gamma_\text{DL-NLO},N)
&= \tilde\chi_0(0,N) + \as \gammaa_0 \tilde\chi_0'(0,N)
  + \as^2 \[\gammaa_1 \tilde\chi_0'(0,N) + \frac12 \gammaa_0^2 \tilde\chi_0''(0,N)\]\nonumber\\ &\quad+ \Ord(\as^3),\\
\tilde\chi_1(\gamma_\text{DL-NLO},N)
&= \tilde\chi_1(0,N) + \as \gammaa_0 \tilde\chi_1'(0,N) + \Ord(\as^2),\\
\chi_1^{\rm corr}(\gamma_\text{DL-NLO},N,\as)
&= \chi_1^{\rm corr}(0,N,0) + \as \[\gammaa_0 \partial_M\chi_1^{\rm corr}(0,N,0)+\partial_{\as}\chi_1^{\rm corr}(0,N,0)\] \nonumber\\ &\quad+ \Ord(\as^2),
\end{align}
where implicit derivatives denoted with a $\prime$ are with respect to $M$, i.e.\ the first argument.
The new functions appearing in the equations above are given by
\begin{subequations}\label{eq:chi00+}
\begin{align}
\frac12 \tilde\chi_0''(0,N) &= -\frac{\chi_{01}}{(1+N)^3} + \chi_{01}\[\zeta_3-\frac12\psi_2(1+N)\], \\
\tilde\chi_1(0,N) &= \tilde\chi_1^{\rm u}(0,0) \nonumber \\
&= \chi_{01}^2\[\frac52\zeta_3-\frac1{24}\] + \chi_{01}\beta_0\zeta_2 + \chi_{02}\[\frac{53}{18}-\zeta_2\]-\chi_{11}, \\
\tilde\chi_1'(0,N) &= \tilde\chi_1^{\rm u\prime}(0,N) \nonumber \\
&=
\(\frac{3\chi_{01}^2}{4}+\chi_{02}\)\(\frac1{(3+2N)^2}+\frac1{(1-2N)^2}\)
\[\frac{9\zeta_2}{32}-\frac3{128}\(\Phi_1(N)-16+2\psi_1(\tfrac14)\)\]
\nonumber\\ &\quad
+\(\frac{\chi_{01}^2}{4}+11\chi_{02}\)\frac2{(1+2N)^2}
\[-\frac{9\zeta_2}{32}-\frac3{128}\(\Phi_1(N)-2\psi_1(\tfrac14)\)\]
\nonumber\\ &\quad
+\[\(\frac{9\chi_{01}^2}{2048}+\frac{3\chi_{02}}{512}\)\(\frac1{3+2N}-\frac1{1-2N}\)
+\(\frac{3\chi_{01}^2}{1024}+\frac{33\chi_{02}}{256}\)\frac1{1+2N}\]\Phi_2(N)
\nonumber\\ &\quad
+\chi_{11}\[\psi_2(2+N)-\zeta_2\]
-2\beta_2\chi_{01} \zeta_3
\nonumber\\ &\quad
+\chi_{02}\[\frac{887}{108}+\frac{167}{24}\zeta_2+\frac32\zeta_3-\frac2{(1+N)^3}
-\frac{13}6\psi_1(1+N)-\frac{47}{48}\psi_1(\tfrac14)\]
\nonumber\\ &\quad
+\chi_{01}^2\bigg[\frac{23}{144}-\frac{\zeta_2}{32}+\frac{37}{40}\zeta_2^2+\frac{\zeta_2}8\Phi_1(N)
+\frac1{64}\psi_1(\tfrac14) +\(\frac{\zeta_2}2- \frac18\)\psi_1(1+N)
\nonumber\\ &\qquad\qquad
+\frac12\psi_1^2(1+N) +\frac1{12}\psi_3(1+N) - \phi_L^{+\prime}(1+N)\bigg]
, \\
\Phi_n(N) &= \psi_n\(\frac{1+N}{2}\) - \psi_n\(1+\frac{N}{2}\),\\
\chi_1^{\rm corr}(0,N,0) &= \beta_0\chi_{01} \[\psi_1(1+N)-2\zeta_2\], \\
\partial_M\chi_1^{\rm corr}(0,N,0) &= \beta_0\chi_{01}\[3\zeta_3-\frac12\psi_2(1+N)\],\\
\partial_{\as}\chi_1^{\rm corr}(0,N,0) &= \beta_0\frac{2\chi_{02}}{(1+N)^3}.
\end{align}
\end{subequations}
Plugging each expansion in Eq.~\eqref{eq:duality} we find
\begin{align}
N &= N +\as\[ \frac{\chi_{01}}{1+N} + \tilde\chi_0(0,N) - \(\frac{\chi_{01}}{2} + \tilde\chi_0(0,1)\)f_{\rm mom}(N)\] \nonumber\\
&\qquad\;+ \as^2 \bigg[
\frac{\gammat_2}{\gammaa_0'}
+\frac{\chi_{02}+\chi_{01}\gammaa_0}{(1+N)^2} + \frac{\chi_{11}}{1+N}
+\gammaa_0 \tilde\chi_0'(0,N) +\tilde\chi_1(0,N) +\chi_1^{\rm corr}(0,N,0)\nonumber\\
&\qquad\qquad\quad-\(\frac{\chi_{02}}{4} +\frac{\chi_{11}}{2}+ \tilde\chi_1(0,1) +\chi_1^{\rm corr}(0,1,0) \)f_{\rm mom}(N)
\bigg]\nonumber\\
&\qquad\;+ \as^3 \bigg[
\frac{\gammat_3}{\gammaa_0'}-\frac{\gammat_2\gammaa_1'}{\gammaa_0'^2}
+\frac{\chi_{03}+2\chi_{02}\gammaa_0+\chi_{01}\gammaa_0^2}{(1+N)^3}
  + \frac{\chi_{12}+\chi_{11}\gammaa_0+\chi_{01}\gammaa_1}{(1+N)^2} + \frac{\chi_{21}}{1+N}\nonumber\\
&\qquad\qquad\quad +\gammaa_1 \tilde\chi_0'(0,N) + \frac12 \gammaa_0^2 \tilde\chi_0''(0,N)
+ \gammaa_0 \tilde\chi_1'(0,N) + \gammaa_0 \partial_M\chi_1^{\rm corr}(0,N,0)+ \partial_{\as}\chi_1^{\rm corr}(0,N,0)
\nonumber\\
&\qquad\qquad\quad-\( \frac{\chi_{03}}{8} + \frac{\chi_{12}}{4} + \frac{\chi_{21}}{2} + \partial_{\as}\chi_1^{\rm corr}(0,1,0) \)f_{\rm mom}(N)
\bigg]\nonumber\\
&\qquad\;+ \Ord(\as^4).
\end{align}
Using Eq.~\eqref{eq:chi00}, we find that both the $\Ord(\as^0)$ and $\Ord(\as)$ contributions vanish automatically.
From the $\Ord(\as^2)$ and $\Ord(\as^3)$ terms it immediately follows
\begin{align}\label{eq:gamma2}
\gammat_2(N) &= -\gammaa_0'
\bigg[ \frac{\chi_{02}}{(1+N)^2} +\frac{\chi_{11}}{1+N} + \tilde\chi_1(0,N) + \chi_{1}^{\rm corr}(0,N,0)
                    + \chi_{01}\[\psi_1(1+N)-\zeta_2\]\gammaa_0 \nonumber\\
  &\qquad\qquad -\(\frac{\chi_{02}}4 + \frac{\chi_{11}}2 + \tilde\chi_1(0,1) +\chi_1^{\rm corr}(0,1,0)\)f_{\rm mom}(N) \bigg].\\
\gammat_3(N) &= \frac{\gammat_2\gammaa_1'}{\gammaa_0'} \nonumber\\
&\quad-\gammaa_0' \bigg[
\frac{\chi_{03}+2\chi_{02}(\gammaa_0+\beta_0)}{(1+N)^3}
  + \frac{\chi_{12}+\chi_{11}\gammaa_0}{(1+N)^2} + \frac{\chi_{21}}{1+N}
+ \chi_{01}\[\zeta_3-\frac12\psi_2(1+N)\]\gammaa_0^2
\nonumber\\ 
&\qquad\qquad
+ \chi_{01}\[\psi_1(1+N)-\zeta_2\]\gammaa_1
+ \gammaa_0 \tilde\chi_1'(0,N)+ \beta_0\chi_{01}\[3\zeta_3-\frac12\psi_2(1+N)\]\gammaa_0
\nonumber\\ 
&\qquad\qquad-\( \frac{\chi_{03}+2\beta_0\chi_{02}}{8} + \frac{\chi_{12}}{4} + \frac{\chi_{21}}{2} \)f_{\rm mom}(N) \bigg],
\label{eq:gamma3}
\end{align}
where we have partially used information from Eqs.~\eqref{eq:chi00} and \eqref{eq:chi00+}.
Further using Eq.~\eqref{eq:chi00+} we can rewrite Eq.~\eqref{eq:gamma2} as\footnote
{Note that this expression differs from the analogous in Ref.~\cite{Bonvini:2017ogt} by NNLL terms,
  due to the correction to the function $\chi_1^{\rm corr}$.}
\begin{align}\label{eq:gamma2v2}
  \gammat_2(N) &= -\gammaa_0'
  \bigg[ \frac{\chi_{02}}{(1+N)^2} +\frac{\chi_{11}}{1+N} + \rho
  + \chi_{01}\[\psi_1(1+N)-\zeta_2\](\gammaa_0+\beta_0) \nonumber\\
 &\qquad\qquad -\(\frac{\chi_{02}}4 + \frac{\chi_{11}}2 + \rho-\beta_0\chi_{01}\)f_{\rm mom}(N) \bigg]
\end{align}
with
\begin{align}\label{eq:rho}
  \rho &= \chi_{01}^2\[\frac52\zeta_3-\frac1{24}\] + \chi_{02}\[\frac{53}{18}-\zeta_2\]-\chi_{11} \nonumber\\
       &= \frac{1}{\pi^2} \[ C_A^2\(-\frac{74}{27}+\frac{11}{12}\zeta_2+\frac52\zeta_3\)
         + n_fC_A\(\frac4{27}+\frac16\zeta_2\)
         + n_fC_F\(\frac7{27}-\frac13\zeta_2\)
         \] .
\end{align}
The N$^3$LO function $\gammat_3(N)$ cannot be simplified significantly, so we do not manipulate it further.

We now move to the running-coupling contributions.
The RC correction to the duality is implemented through the function
\begin{align}\label{eq:A.16}
\Delta\gamma_{ss}^{\rm rc}(N,\as)
&= -\beta_0\as \[ \frac{\chi_0''(M)\chi_0(M)}{2{\chi_0'}^2(M)} - 1\]_{M=\gamma_s(\as/N)}
\nonumber\\ &
= -\as^4\beta_0\frac{\chi_{01}^3}{N^3}12\zeta_3 + \Ord(\as^5),
\end{align}
which contributes at NLL. Further RC corrections from RC resummation, Eq.~\eqref{eq:RCsol}, are NNLL corrections.
They amount to
\begin{align}\label{eq:DeltaRCNLO}
\Delta_\text{DL-NLO}\gamma_{\rm rc}(N,\as)
&\equiv \gamma_{\rm rc}(N,\as) - \Bigg[M_{\rm min} - \sqrt{\frac{N-c}{\kappa/2 +(N-c)/M_{\rm min}^2}} - \beta_0\as \nonumber\\
&\qquad\qquad\qquad\quad+\frac14 \beta_0\as^2\(3\frac{\kappa'-2c'/M_{\rm min}^2}{\kappa+2(N-c)/M_{\rm min}^2}-\frac{c'}{N-c}\)\Bigg]
\nonumber\\
&= \beta_0^2\as^3 \frac{\kappa_0}{16N} \nonumber\\
&\quad+ \beta_0^2\as^4 \[
\frac{\kappa_0(192 c_0-7\kappa_0)}{1024N^2} +
\frac{\kappa_0(2m_1^{\rm NLO}-3\beta_0) +T'\kappa_1^{\rm NLO}}{16N}
\]\nonumber\\
&\quad+ \Ord(\as^5),
\end{align}
where as before $T'$ is either $2$ or $1$ depending on the approximate $\as$ dependence chosen.
To cancel the singularity mismatch between Eq.~\eqref{eq:DeltaRCNLO} and Eq.~\eqref{eq:A.16}
we further need the matching function
\begin{align}\label{eq:gammassmatch}
\gamma^{ss}_{\rm match}(N,\as)
&= \frac14\beta_0\as^2 \[ \frac{c_0}{N-\as c_0} - \frac{c'}{N-c} +\frac{c'-c_0}{N} \]\nonumber\\
&= -\as^4\beta_0c_0\frac{(1+T') c_1^{\rm NLO}}{4N^2} + \Ord(\as^5).
\end{align}
The coefficients above are the ones obtained from the NLO kernel,
given in Sect.~\ref{sec:coefficients}.

Putting everything together according to Eq.~\eqref{eq:gammaresNLLdef}, we obtain
\begin{align}\label{eq:gammaNLLexp}
\gamma^{\rm res\, NLL}(N,\as)
&= \as \gammaa_0(N) + \as^2 \gammaa_1(N) \nonumber\\
&+ \as^3 \bigg\{ \beta_0^2 \frac{\kappa_0}{16} \(\frac1N - f_{\rm mom}(N)\) \nonumber\\
&\qquad\quad-\gammaa_0' \bigg[ \frac{\chi_{02}}{(1+N)^2} +\frac{\chi_{11}}{1+N} + \rho
  + \chi_{01}\[\psi_1(1+N)-\zeta_2\](\gammaa_0+\beta_0)\nonumber\\
&\qquad\qquad\qquad-\(\frac{\chi_{02}}4 + \frac{\chi_{11}}2 + \rho-\beta_0\chi_{01}\)f_{\rm mom}(N) \bigg]\bigg\} \nonumber\\
&+ \as^4 \bigg\{\frac{\rho_3}{N^3} +\frac{\rho_2}{N^2} +\frac{\rho_1}{N} - (\rho_3+\rho_2+\rho_1) f_{\rm mom}(N) \nonumber\\
&\qquad\quad-\gammaa_1' \bigg[ \frac{\chi_{02}}{(1+N)^2} +\frac{\chi_{11}}{1+N} + \rho
  + \chi_{01}\[\psi_1(1+N)-\zeta_2\](\gammaa_0+\beta_0)\nonumber\\
&\qquad\qquad\qquad-\(\frac{\chi_{02}}4 + \frac{\chi_{11}}2 + \rho-\beta_0\chi_{01}\)f_{\rm mom}(N) \bigg]\nonumber\\
&\qquad\quad-\gammaa_0' \bigg[
\frac{\chi_{03}+2\chi_{02}(\gammaa_0+\beta_0)}{(1+N)^3}
  + \frac{\chi_{12}+\chi_{11}\gammaa_0}{(1+N)^2} + \frac{\chi_{21}}{1+N}
\nonumber\\ 
&\qquad\qquad\qquad
-\( \frac{\chi_{03}+2\beta_0\chi_{02}}{8} + \frac{\chi_{12}}{4} + \frac{\chi_{21}}{2} \)f_{\rm mom}(N)
\nonumber\\ 
&\qquad\qquad\qquad
+ \chi_{01}\[\zeta_3-\frac12\psi_2(1+N)\]\gammaa_0 (\gammaa_0+\beta_0)
+ \chi_{01}2\zeta_3\beta_0\gammaa_0
\nonumber\\ 
&\qquad\qquad\qquad
+ \chi_{01}\[\psi_1(1+N)-\zeta_2\]\gammaa_1
+ \gammaa_0 \tilde\chi_1'(0,N)
\bigg]\bigg\}\nonumber\\
&+\Ord(\as^5),
\end{align}
where
\begin{subequations}
\begin{align}
\rho_3 &= -\beta_0\chi_{01}^3 12\zeta_3, \\
\rho_2 &=  \beta_0^2\kappa_0\[\frac3{16}c_0-\frac7{1024}\kappa_0\] - \beta_0c_0c_1^{\rm NLO}\frac{1+T'}{4}, \\
\rho_1 &=  \beta_0^2 \frac{\kappa_0(2m_1^{\rm NLO}-3\beta_0) +T'\kappa_1^{\rm NLO}}{16},
\end{align}
\end{subequations}
and with $\tilde\chi_1'(0,N)$ given in Eq.~\eqref{eq:chi00+}.
The $\Ord(\as^4)$ contribution in Eq.~\eqref{eq:gammaNLLexp} is a new result.

Eq.~\eqref{eq:gammaNLLexp} is rather complex and not optimal to be used in a numerical code.
In particular, we also need to compute the inverse Mellin transform of this object
in order to provide the expansion of the resummed splitting functions, and it is clearly
very complicated (if not impossible) to obtain analytic expressions for such inverse.
Thus, we consider a pole expansion of the $\Ord(\as^4)$ term of Eq.~\eqref{eq:gammaNLLexp}, $\gamma_3^{\rm NLL}$.
Specifically, we compute the residues of all the poles in $N=0,-1,-2$,
and construct an approximation based only on these terms,
\beq
\gamma_3^{\rm NLL}(N) \simeq 
\sum_{j=0}^3 \frac{g_{0j}}{N^{j+1}} +
\sum_{j=0}^6 \frac{g_{1j}}{(N+1)^{j+1}} +
\sum_{j=0}^2 \frac{g_{2j}}{(N+2)^{j+1}}.
\eeq
This approximation has the advantage of describing in $x$ space all contributions behaving
as $x^k \log^jx$ for all non-vanishing terms with $j\geq0$ and for $k=-1,0,1$,
namely all non-vanishing terms at small $x$ plus the leading corrections to them.
Of course this approximation will not be accurate at large $x$, but since the final results
will be damped at large $x$ the inaccuracy should be negligible.
To verify the quality of our approximation, we have compared it with
a simpler approximation which does not include the contributions from the poles in $N=-2$,
i.e.\ those terms behaving as $x\log^jx$ in $x$ space.
The difference between the two results is almost imperceptible.

Before concluding, it is useful to extract from Eq.~\eqref{eq:gammaNLLexp} its small-$x$ behaviour up to NLL,
which provides a prediction for the yet unknown four-loop anomalous dimensions.
Expanding the anomalous dimensions Eq.~\eqref{eq:gammaapprox} in $N=0$,
\begin{align}
\gammaa_0(N) &= \frac{a_{11}}{N} + a_{10} + \Ord(N), &
\gammaa_0'(N) &= -\frac{a_{11}}{N^2} + \Ord(N^{0}), \\
\gammaa_1(N) &= \frac{a_{21}}{N} + \Ord(N^{0}), &
\gammaa_1'(N) &= -\frac{a_{21}}{N^2} + \Ord(N^{0}),
\end{align}
and knowing that $\tilde\chi_1'(0,N)$ is finite in $N=0$, we can easily compute the
$N=0$ expansion of Eq.~\eqref{eq:gammaNLLexp}.
For this result, we also need the function $\phi_L^+(M)$ and its derivative evaluated in $M=0,1$,\footnote
{We could not compute the derivatives analytically, so we have used the PSLQ algorithm to find the results of the integrals.}
\begin{align}
\phi_L^{+}(0) &= -\zeta_2\log2+\frac{13}8\zeta_3, &
\phi_L^{+}(1) &= \zeta_2\log2-\frac58\zeta_3, \\
\phi_L^{+\prime}(0) &= -\frac{13}{16}\zeta_4, &
\phi_L^{+\prime}(1) &= -\frac{3}{16}\zeta_4.
\end{align}
Denoting with $\gamma_+^{(n)}$ the exact anomalous dimensions at $\Ord(\as^{n+1})$, we find
\begin{align}
\gamma_+^{(2)}(N)
  &=\frac{C_A}{\pi^3N^2} \frac{C_A^2 (54 \zeta_3 + 99 \zeta_2 - 395) + n_f (C_A - 2 C_F) (18 \zeta_2 - 71)}{108}+\Ord\(\frac1N\),\\
\gamma_+^{(3)}(N)
    &= \frac{C_A^4}{\pi^4N^4}2\zeta_3 \nonumber\\
  &\quad+ \frac{C_A^2}{\pi^4N^3}
    \bigg[C_A^2\(-\frac{1205}{162} + \frac{67}{36}\zeta_2 + \frac14 \zeta_2^2 - \frac{77}{6} \zeta_3\)
    + n_fC_A\(-\frac{233}{162} + \frac{13}{36}\zeta_2 + \zeta_3\)\nonumber\\
  &\qquad\qquad\quad+ n_fC_F\(\frac{233}{81} - \frac{13}{18}\zeta_2 + \frac43 \zeta_3\)
    \bigg]\nonumber\\
  &\quad+\Ord\(\frac1{N^2}\)
\end{align}
To our knowledge, the NLL term at N$^3$LO was never explicitly presented in the literature.
Note that the scheme is $Q_0\MSbar$.
The conversion to $\MSbar$ was presented in Sect.~\ref{sec:MSbar}.

\subsection{Computation of the coefficients describing the minimum of the BFKL kernel}
\label{sec:coefficients}

In this section we compute the expansion coefficients of $c(\as)$ and $\kappa(\as)$ of the BFKL kernel in symmetric variables.
We will assume that the minimum is not in $M=1/2$, but in
\beq\label{eq:Mminexp}
\tilde M_{\rm min}(\as) = \frac12 + \as \tilde m_1+\as^2\tilde m_2+....
\eeq
The expansion coefficients of the minimum can be found by solving perturbatively the minimum condition
\beq\label{eq:Mmincond}
\left.\partial_M\chi(M,\as)\right|_{M=\tilde M_{\rm min}(\as)} = 0,
\eeq
where this $\chi(M,\as)$ is the on-shell kernel in symmetric variables, obtained by the on-shell condition
\beq\label{eq:onshellcond}
\chi(M,\as) = \bc\(M,\chi(M,\as),\as\),
\eeq
where $\bc(M,N,\as)$ is the off-shell kernel in symmetric variables, related to the DL kernel via~\cite{Altarelli:2005ni,Bonvini:2017ogt}
\beq
\bc(M,N,\as) = \chi_\Sigma(M+N/2,N,\as).
\eeq
Using Eq.~\eqref{eq:onshellcond} in Eq.~\eqref{eq:Mmincond} we get
(here $M$-partial derivatives act on the first argument, which is then set to the written value)
\beq\label{eq:Mmineq}
\partial_M\bc\(\tilde M_{\rm min}(\as),c(\as),\as\) = 0,
\eeq
having used the definition $c(\as) = \chi(\tilde M_{\rm min}(\as),\as)$ which, in terms of the off-shell kernel, leads to the implicit equation
\beq\label{eq:ceq}
c(\as) = \bc(\tilde M_{\rm min}(\as),c(\as),\as).
\eeq
Equations \eqref{eq:Mmineq} and \eqref{eq:ceq} can be iteratively solved order by order in perturbation theory.
Introducing the expansions
\begin{align}
  c(\as) &= \as c_0 + \as^2 c_1 + ... \label{eq:cexp}\\
  \bc(M,N,\as) &= \as \bc_0(M,N) + \as^2 \bc_1(M,N) + ...
\end{align}
we have
\begin{align}
  &0 = \as \partial_M\bc_0\(\frac12 + \as \tilde m_1+..., \as c_0 + ...\)
   + \as^2 \partial_M\bc_1\(\frac12 + \as \tilde m_1+..., \as c_0 + ...\) +... \\
  &\as c_0 + \as^2 c_1 + ...
  = \as \bc_0\(\frac12 + \as \tilde m_1 +..., \as c_0 + ...\)
  + \as^2 \bc_1\(\frac12 + \as \tilde m_1+..., \as c_0 + ...\) +...
\end{align}
and further expanding we get
\begin{align}
  &0 = \as \partial_M\bc_0\(\frac12,0\) + \as^2\[\tilde m_1 \partial_M^2\bc_0\(\frac12,0\) + c_0 \partial_N \partial_M\bc_0\(\frac12,0\) + \partial_M\bc_1\(\frac12,0\) \]+... \\
  &\as c_0 + \as^2 c_1 + ...
  = \as \bc_0\(\frac12,0\) + \as^2 \[\tilde m_1 \partial_M\bc_0\(\frac12,0\) + c_0 \partial_N\bc_0\(\frac12,0\) +\bc_1\(\frac12,0\) \]+...
\end{align}
From these equations we find that $\partial_M\bc_0\(\frac12,0\)=0$, which was our assumption in Eq.~\eqref{eq:Mminexp}.
Note that, more in general, we have $\partial_M\bc_0\(\frac12,N\)=0$ for any $N$, which implies that $\partial_N\partial_M\bc_0\(\frac12,0\)=0$.
Before listing the results, let us also consider the curvature $\kappa$, with expansion
\beq\label{eq:kappaexp}
\kappa(\as) = \as \kappa_0 + \as^2\kappa_1 + ...
\eeq
which is given by $\kappa(\as) = \partial_M^2\chi(\tilde M_{\rm min},\as)$ and hence,
in terms of off-shell kernel~\cite{Bonvini:2012sh},
\beq
\kappa(\as) = \left.\frac{\partial_M^2\bc}{1-\partial_N\bc}\right|_{M=\tilde M_{\rm min}(\as), N=c(\as)}.
\eeq
Expanding this equation we find
\begin{align}
  \kappa(\as)
  &= \[\as\partial_M^2\bc_0\(\frac12 + \as \tilde m_1 +..., \as c_0 + ...\) + \as^2\partial_M^2\bc_1\(\frac12 + ..., 0 + ...\) + ...\]\nonumber\\
  &\qquad \times\[1+\as\partial_N\bc_0\(\frac12 + ..., 0 + ...\)+...\] \nonumber\\
  &=\as\partial_M^2\bc_0\(\frac12,0\) \nonumber\\
  &\quad + \as^2
    \[\partial_M^2\bc_1\(\frac12,0\) + \partial_M^2\bc_0\(\frac12,0\) \partial_N\bc_0\(\frac12,0\) + c_0\partial_N \partial_M^2\bc_0\(\frac12,0\) + \tilde m_1\partial_M^3\bc_0\(\frac12,0\)\] \nonumber\\
&\quad +...
\end{align}
Note that the last term vanishes due to the fact that the LO kernel is symmetric, and hence all its odd $M$-derivatives are zero.
Putting everything together, we then obtain
\begin{align}
c_0 &= \bc_0,\\
\kappa_0 &= \partial_M^2\bc_0,\\
\tilde m_1 &= -\frac{\partial_M\bc_1}{\kappa_0}, \\
c_1 &= \bc_1 + c_0 \partial_N\bc_0, \\
\kappa_1 &= \partial_M^2\bc_1 + \kappa_0 \partial_N\bc_0 + c_0 \partial_N\partial_M^2\bc_0,
\end{align}
where every off-shell kernel is implicitly assumed to be computed in $M=1/2$, $N=0$.
Explicitly, we have at lowest order
\begin{subequations}
\begin{align}
c_0 &= \chi_{01}4\log2,\\
\kappa_0 &= \chi_{01} 28\zeta_3,
\end{align}
\end{subequations}
while for the higher order coefficients the result depends on the actual kernel considered.
For DL-LO kernel we have
\begin{subequations}
\begin{align}
\tilde m_1^{\rm LO} &= 0, \\
c_1^{\rm LO} &= 8\chi_{02} - 8\chi_{01}^2\zeta_2\log2 \nonumber\\
&= -15.00496429 - 0.04503163717 n_f, \\
\kappa_1^{\rm LO} &= 192\chi_{02} - \chi_{01}^2\(144\zeta_2^2\log2+56\zeta_2\zeta_3\) \nonumber\\
&= -507.744719 - 1.080759292 n_f,
\end{align}
\end{subequations}
while for DL-NLO kernel we have
\begin{subequations}
\begin{align}
  \tilde m_1^{\rm NLO}
  &= \beta_0\(\frac{4\log2}{7\zeta_3} - \frac12 \), \\
  c_1^{\rm NLO}
  &= \chi_{02}\[\frac{33\pi^3}{64}-\frac{26}{3}\log2\] + 8\beta_0\chi_{01}\[\log2-\log^22\]
    +\chi_{11}4\log2\nonumber\\
  &\quad+ \chi_{01}^2\[\frac{73\pi^3}{786}-\frac{\log2}2-2\zeta_2\log2-\frac{11}2\zeta_3+2\phi_L^+\(\frac12\)\] \nonumber \\
  &= -11.696833425 - 0.4102968810 n_f, \\
  \kappa_1^{\rm NLO}
  &= \chi_{02}\[-\frac{3\pi^3}{32}+\frac{55\pi^5}{64}-\frac{182}{3}\zeta_3\]
    +\chi_{11}28\zeta_3
    + \beta_0\chi_{01}\[64\log2+56\zeta_3-112\log2\zeta_3\]
    \nonumber\\
  &\quad+ \chi_{01}^2\[-\frac{9\pi^3}{128}+\frac{79\pi^5}{786}-\frac72\zeta_3-14\zeta_2\zeta_3-372\zeta_5+2\phi_L^{+\prime\prime}\(\frac12\)\] \nonumber \\
&= -494.250393369 - 5.23585215538 n_f.
\end{align}
\end{subequations}
Unfortunately, we could not be able to express the function $\phi_L^+(M)$, Eq.~\eqref{eq:phiL}, and its second derivative
computed in $M=1/2$ in terms of elementary constants.
Finally, the differences of the coefficients needed for the LL$^\prime$ result have the following numerical values:
\begin{subequations}
\begin{align}
\delta m_1 &= -0.1492429211 + 0.00904502552 n_f,\label{eq:deltam1}\\
\delta c_1 &= 3.308130862 - 0.3652652438 n_f, \\
\delta\kappa_1 &= 13.49432608 - 4.155092863 n_f.
\end{align}
\end{subequations}
Note that ultimately one wants to construct the anomalous dimensions which are dual
to the BFKL kernel in asymmetric (or DIS) variables.
In Ref.~\cite{Bonvini:2017ogt} we argued that $\Delta_\text{DL-(N)LO}\gamma_{\rm rc}^{\rm (N)LL}$,
which contains the RC resummation corrections to be added to the DL result,
could be computed directly from the kernel in symmetric variables,
as the conversion from symmetric to asymmetric amounts to adding $+N/2$ to the anomalous dimension,
which is then subtracted again by the matching.
However, the argument was superficial, and induced by the collinear approximation of the kernel used for computing the resummation
of RC effects. Indeed, the actual DL BFKL kernel in symmetric variables behaves asymptotically as $-2M$ in the collinear region,
which by duality (either fixed-coupling or running-coupling) reproduces the $-N/2$ behaviour which is then removed by the conversion
from symmetric to asymmetric variables.
Therefore, using a collinear approximation to the kernel is more appropriate if the kernel is the one in asymmetric variables.
Thus, the RC parameters ($c$, $\kappa$ and $M_{\rm min}$) should be the ones of such kernel.
While the curvature and the value of the kernel at the minimum are not sensitive to the conversion, the position of the minimum is,
and thus in the \texttt{3.0} version of the \texttt{HELL} code we use the results of Ref.~\cite{Bonvini:2017ogt}
with 
\beq\label{eq:newMmin}
M_{\rm min}(\as)= \tilde M_{\rm min}(\as)+c(\as)/2.
\eeq
Note that in the expansion of this appendix this implies that $m_1=\tilde m_1+c_0/2$,
while $\delta m_1$, Eq.~\eqref{eq:deltam1}, remains unchanged.

\section{Running coupling corrections to the DGLAP-BFKL duality}
\label{sec:bug}

The BFKL kernel (in symmetric variables) is not fully symmetric for the exchange $M\leftrightarrow1-M$ because of a number of effects,
all induced by the running of the strong coupling.
The origin of these effects can be traced back to (see e.g.~\cite{Altarelli:2005ni}):
\begin{itemize}
\item the scale at which $\as$ is computed in different pieces of the fixed-order kernel;
\item the conversion from the unintegrated ($\kt$-dependent) PDFs, to which the original BFKL equation refers,
  to the integrated (collinear) PDFs, to which the DGLAP equation refers;
\item the so-called running coupling corrections to the duality,
  namely the correct derivation of the duality between DGLAP and BFKL taking into account the scale dependence of the strong coupling.
\end{itemize}
These effects have been extensively discussed in the literature and they will not be rederived;
here we limit ourselves to present them.
All these effects can be implemented as corrections to the BFKL kernel,
which then gives by naive (i.e., fixed-coupling) duality the correct NLL contributions to the resummed anomalous dimension.
These corrections amount to the following expression to be added to the symmetric kernel,
\beq\label{eq:Deltachi}
\Delta\chi(M,\as) =
\as^2\beta_0 \[ -\psi_1(1-M) + \frac{\chi_0(M)}{M} + \frac{\chi_0(M)\chi_0''(M)}{2\chi_0'(M)} \] + \Ord(\as^3),
\eeq
where the three contributions in square brackets correspond to the three items above, respectively.
The last term contains a $\chi_0'(M)$ in the denominator, which is singular in $M=1/2$,
and thus produces a new singularity that makes the result perturbatively unstable.
It has been realised long ago~\cite{Altarelli:2003hk,Altarelli:2005ni} that these singular contributions can be
removed (resummed) if the running coupling corrections to duality are accounted for
to all orders rather than included perturbatively.
This is the goal of the RC resummation, which provides an anomalous dimension where these contributions
are resummed by solving the BFKL equation with running coupling (in a given approximation).
Thus, it is convenient to translate the last correction to the kernel in a correction to the anomalous dimension,
which is then regulated when the RC resummed result is added.
This is always possible, and a simple computation reveals that one can pour the last term of Eq.~\eqref{eq:Deltachi}
into
\beq\label{eq:Deltagamma}
\Delta\gamma(N,\as) = -\as\beta_0 \[\frac{\chi_0(M)\chi_0''(M)}{2\chi_0'^2(M)}\]_{M=\gamma_s(\as/N)},
\eeq
up to subleading logarithms.
This expression, however, is not yet optimal for numerical implementation.
Indeed, expanding this contribution we get
\beq\label{eq:Deltagammaas}
\Delta\gamma(N,\as) = -\as\beta_0 +\Ord(\as^4).
\eeq
This implies that the naive dual of the kernel
(the symmetric kernel plus $\Delta\chi$, Eq.~\eqref{eq:Deltachi}, without the last term)
does not coincide with the high-energy limit of the exact anomalous dimension at $\Ord(\as)$.
This is problematic when one wants to resum the collinear singularities
of the kernel using $\chi_s$, the dual of the LO anomalous dimension, as the singularities would not cancel.
One could in principle cure this problem by computing $\chi_s$ from a modified fixed-order anomalous dimension
to which the $-\as\beta_0$ term, Eq.~\eqref{eq:Deltagammaas}, has been subtracted, namely $\gammaa_0\to\gammaa_0+\as\beta_0$.
However, this becomes too artificial and not very physical.
A better way to solve the issue is to pour back the lowest order term of the expansion of $\Delta\gamma$ into $\Delta\chi$.
We would thus have (using the same names for the modified objects)
\begin{align}
\Delta\gamma(N,\as) &= -\as\beta_0 \[\frac{\chi_0(M)\chi_0''(M)}{2\chi_0'^2(M)}-1\]_{M=\gamma_s(\as/N)},\label{eq:Deltagammanew}\\
\Delta\chi(M,\as) &=
\as^2\beta_0 \[ -\psi_1(1-M) + \frac{\chi_0(M)}{M} + \chi_0'(M) \] + \Ord(\as^3). \label{eq:Deltachinew}
\end{align}
In this way, the singularities of the kernel can be resummed with the standard $\chi_s$,
and the dual anomalous dimension receives a correction of order $\Ord(\as^4)$,
which is then accounted for to all orders by RC resummation.
This $\Delta\gamma$, Eq.~\eqref{eq:Deltagammanew}, is indeed the function $\Delta\gamma_{ss}^{\rm rc}$, Eq.~\eqref{eq:A.16},
used in our construction.
The $\Delta\chi$ contribution, instead, generates the $\chi_1^{\rm corr}$ term appearing in the off-shell kernel, Eq.~\eqref{eq:chiDLNLO}.
Indeed, the naive off-shell extension of $\Delta\chi$ gives
\beq  \label{eq:Deltachinewoff}
\Delta\chi(M,\as) \to
\as^2\beta_0 \[ -\psi_1(1-M+N) + \frac{\chi_0(M,N)}{M} + \chi_0'(M,N) \] + \Ord(\as^3).
\eeq
The actual $\chi_1^{\rm corr}$ that we use, Eq.~\eqref{eq:chi1corr},
further adds a higher order contribution to resum the singularity in $M=1+N$,
thus stabilizing the resulting kernel without affecting the logarithmic accuracy of the resummed anomalous dimension.

In our previous implementation of the resummation, Ref.~\cite{Bonvini:2017ogt},
we used a different (wrong) form of $\chi_1^{\rm corr}$, taken from Ref.~\cite{Altarelli:2005ni},
\beq  \label{eq:Deltachinewoffwrong}
\chi_1^{\rm corr,~\tt HELL~2.0}(M,N,\as) =
\as^2\beta_0 \[ -\psi_1(1-M+N) + \frac{\chi_0(M,N)}{M} - \frac{\chi_{01}}{M^2} \] + \Ord(\as^3),
\eeq
which differs from the correct Eq.~\eqref{eq:Deltachinewoff} by the last term.
The singularity of the last term is the correct one, and thus leads to
a proper resummation through $\chi_s$, which is the argument used by Ref.~\cite{Altarelli:2005ni}
to introduce such a ``subtraction'' term. However, the missing $M$-dependent contributions
at $\Ord(\as^2)$ produce spurious NLL contributions in the anomalous dimension.
This NLL effect starts to appear at $\Ord(\as^4)$, which explains why the comparison
of the three-loop anomalous dimension obtained from the resummation to the exact result was successful.
For this reason, the full resummed result after the correction is not very different from the previous bugged one.

\begin{figure}[t]
  \centering
  \includegraphics[width=0.495\textwidth,page=7]{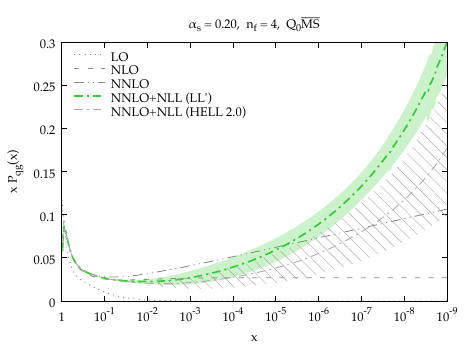}
  \includegraphics[width=0.495\textwidth,page=8]{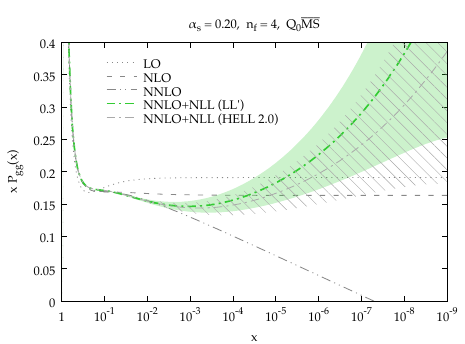}
  \caption{Comparison between resummed and matched splitting functions in \texttt{HELL 3.0} with LL$'$ anomalous dimension
    and \texttt{HELL 2.0}, which contained a bug.}
  \label{fig:bug}
\end{figure}
To conclude, we show in Fig.~\ref{fig:bug} the comparison between the $P_{qg}$ and $P_{gg}$
splitting functions obtained using the LL$'$ anomalous dimension before (dot-dashed gray) and after (dot-dashed green)
the correction of the bug, and the new definition of $M_{\rm min}$, Eq.~\eqref{eq:newMmin}.
Both splitting functions appear to be harder after bug corrections,
which is due to both $\gamma_+^{\rm NLL}$ and $\gamma_+^{\rm LL'}$ being indeed harder.
We stress however that much of this effect is induced by the change in $M_{\rm min}$,
rather than to the correction in $\chi_1^{\rm corr}$.
The uncertainty band on $P_{qg}$ is significantly reduced, again mostly due to the different $M_{\rm min}$ used.
Overall, however, the effect is not dramatic, and likely comparable to (if not smaller than) unknown subleading corrections,
which in $P_{qg}$ are likely underestimated by our uncertainty band, as we already commented in Sect.~\ref{sec:results}.

\phantomsection
\addcontentsline{toc}{section}{References}

\bibliographystyle{jhep}
\bibliography{references}

\end{document}